\documentclass{aa}

\usepackage[varg]{txfonts}

\usepackage{rotating}
\usepackage{xspace}
\usepackage{amssymb}
\usepackage{graphicx}
\usepackage{booktabs}
\usepackage{hyperref}
\usepackage{txfonts}
\usepackage{graphicx, float}
\usepackage{gensymb}
\usepackage{natbib}
\usepackage{epsfig, natbib}
\usepackage{paralist}

\begin{document}

\title
{Evidence for a chemical enrichment coupling of globular clusters and field stars in the Fornax dSph\thanks{This article includes data gathered with the 6.5 meter Magellan Telescopes located at Las Campanas Observatory, Chile.}}

\author{Benjamin Hendricks\inst{\ref{inst1}} \and Corrado Boeche\inst{\ref{inst2}} \and Christian I. Johnson\inst{\ref{inst3}} \and Matthias J. Frank\inst{\ref{inst1}} \and Andreas Koch\inst{\ref{inst1}} \and Mario Mateo\inst{\ref{inst4}} \and John I. Bailey III\inst{\ref{inst4}}}
 
\institute{Zentrum f\"ur Astronomie der Universit\"at Heidelberg, Landessternwarte, K\"onigstuhl 12, 69117, Heidelberg, Germany\label{inst1} \and
Astronomisches Rechen-Institut, Zentrum f\"ur Astronomie der Universit\"at Heidelberg, M\"onchhofstr. 12-14, 69120 Heidelberg, Germany\label{inst2} \and
Harvard-Smithsonian Center for Astrophysics, 60 Garden Street, MS-15, Cambridge, MA 02138, USA\label{inst3} \and
Department of Astronomy, University of Michigan, 1085 South University, Ann Arbor, MI  48109, USA\label{inst4}}

\abstract{The globular cluster H4, located in the center of the Fornax dwarf spheroidal galaxy, is crucial for understanding the formation and chemical evolution of star clusters in low-mass galactic environments. H4 is peculiar because the cluster is significantly more metal-rich than the galaxy's other clusters, is located near the galaxy center, and may also be the youngest cluster in the galaxy. In this study, we present detailed chemical abundances derived from high-resolution (R$\sim28\,000$) spectroscopy of an isolated H4 member star for comparison with a sample of 22 nearby Fornax field stars. We find the H4 member to be depleted in the alpha-elements Si, Ca, and Ti with [Si/Fe]$=-0.35\pm{0.34}$, [Ca/Fe]$=+0.05\pm0.08$, and [Ti/Fe]$=-0.27\pm{0.23}$, resulting in an average [$\mathrm{\alpha}$/Fe]$=-0.19\pm{0.14}$. If this result is representative of the average cluster properties, H4 is the only known system with a low [$\mathrm{\alpha}$/Fe] ratio and a moderately low metallicity embedded in an intact birth environment.
For the field stars we find a clear sequence, seen as an early depletion in [$\mathrm{\alpha}$/Fe] at low metallicities, in good agreement with previous measurements. H4 falls on top of the observed field star [$\mathrm{\alpha}$/Fe] sequence and clearly disagrees with the properties of Milky Way halo stars. We therefore conclude that within a galaxy, the chemical enrichment of globular clusters may be closely linked to the enrichment pattern of the field star population. The low [$\mathrm{\alpha}$/Fe] ratios of H4 and similar metallicity field stars in Fornax give evidence that slow chemical enrichment environments, such as dwarf galaxies, may be the original hosts of alpha-depleted clusters in the halos of the Milky Way and M31.
}


\titlerunning{Chemical Coupling of GCs and Field Stars in the Fornax dSph}
\authorrunning{Hendricks et al.}
\maketitle

\section{Introduction} 
\label{chap_01}

Globular clusters (GCs) are an intriguing class of stellar systems and have been objects of interest for many decades. When resolved, they offer a unique possibility of studying their effectively coeval and mono-metallic stellar populations in detail. In distant galaxies, unresolved globular clusters serve as luminous beacons that can still be analyzed when individual field stars are too faint to observe (e.g., \citealt{Brodie_06}).

While the chemical (self-)enrichment in GCs \emph{after} their formation has been studied in great detail in the recent past (see, e.g., \citealt{Gratton_12} and references therein), relatively little is known about the chemical enrichment of these systems \emph{before} their formation, that is, the chemical evolution of the proto-GC gas embedded in a galactic environment. Specifically, it is not clear whether GCs---despite their undoubtedly unique formation mechanism---reflect the chemical properties of the field stars in the host environment they are born in or whether they instead display distinct chemical enrichment properties.

The alpha-elements (O, Mg, Si, Ca, and Ti) are an important instrument for studying the details of chemical enrichment in stellar systems.
These elements are synthesized almost exclusively in massive stars and therefore trace the contribution ratio between high-mass Supernovae~II (SNe~II; yielding high [$\mathrm{\alpha}$/Fe]) and low-mass Supernova~Ia (SNe~Ia; yielding low [$\mathrm{\alpha}$/Fe]) to the interstellar medium (ISM). 
As a result, the [$\mathrm{\alpha}$/Fe] ratio is enhanced for the first, metal-deficient stars and subsequently drops with increasing [Fe/H] as soon as SNe~Ia start to contribute significantly to the enrichment. The increasing contributions from SNe Ia are seen as an inflection point or \emph{knee} in plots of 
[$\mathrm{\alpha}$/Fe] versus [Fe/H]. Consequently, the evolution of [$\mathrm{\alpha}$/Fe] reveals the extent to which the galaxy can enrich its ISM before SNe~Ia set in, and hence is a measurement for the chemical enrichment efficiency of the system (\citealt{Tinsley_79}, \citealt{Matteucci_90}, \citealt{Lanfranchi_03}). 
Since all stars are effectively coeval in a GC, their $\mathrm{\alpha}$/Fe-ratio represents the chemical enrichment status of the proto-cluster material at the time of formation. 

\begin{figure}[h]
\begin{center}
\includegraphics[width=0.5\textwidth]{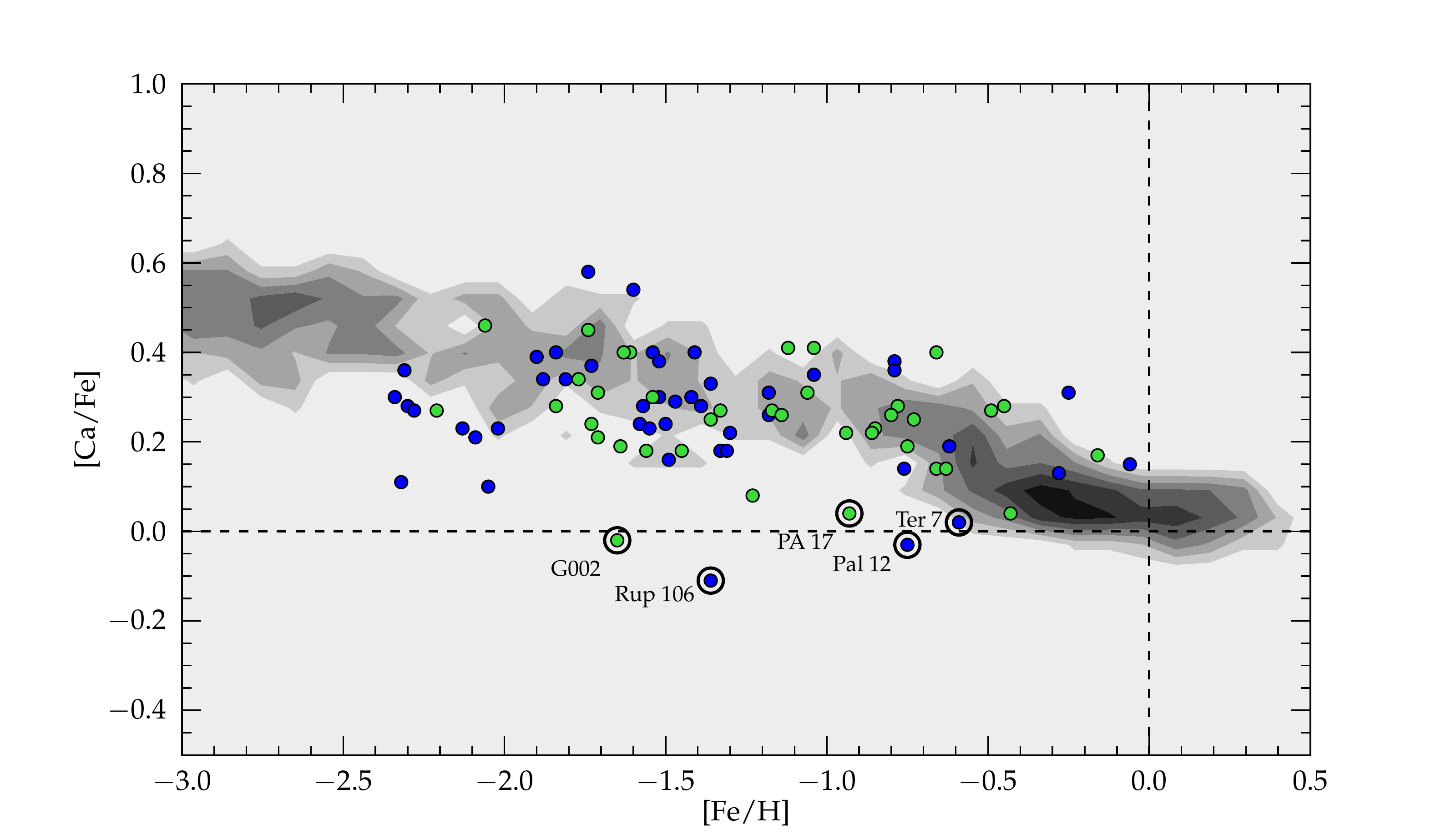}
\caption{Literature compilation of the chemical abundance pattern of Ca as a function of [Fe/H]. Symbols show GCs in the MW from \citet{Pritzl_05} (blue) and M31 (green) from \citet{Colucci_14} and \citet{Sakari_15}. The pattern of MW disc and halo stars are shown as a logarithmically-scaled number density distribution of arbitrary units (data from \citealt{Venn_04} and \citealt{Roederer_14}). Outliers with significantly lower [Ca/Fe] are highlighted with large open black circles.}
\label{figure_intro_1}
\end{center}
\end{figure}

Globular clusters in the Milky Way (MW) halo are found to be almost exclusively alpha-enhanced over the entire observed range of metallicities (\citealt{Pritzl_05}), indicating a fast chemical enrichment of the material from which they formed. Simultaneously, they fall on top of the enhanced [$\mathrm{\alpha}$/Fe] plateau displayed by the field stars (e.g., \citealt{Venn_04}), and therefore no clear distinction between a uniform enrichment scenario and a coupling with the field star properties can be made (see Figure~\ref{figure_intro_1}).
A similar picture is found for the GC system in Andromeda (M31) where a large number of clusters have been studied recently by \citet{Colucci_14} and \citet{Sakari_15} using integrated-light spectroscopy. Although the alpha-evolution of field stars in M31 is not known given the extremely faint magnitudes of individual stars, the [$\mathrm{\alpha}$/Fe] values of the GCs agree well with the field stars and the GCs of the MW. Notably, the clusters on the high-metallicity end in the sample of \citet{Colucci_14} possibly show lower alpha-abundance ratios and may follow the knee of MW field stars, which could be a hint for a common chemical enrichment pattern between field stars and clusters.

Both galaxies host a few, but interesting, outliers with significantly lower [$\mathrm{\alpha}$/Fe] ratios with respect to the field stars at comparable metallicity: Terzan~7, Palomar~12, and the younger system Ruprecht~106 in the MW (\citealt{Sbordone_05}, \citealt{Cohen_04}, \citealt{Brown_97}), G002 and PA17 in M31 (\citealt{Colucci_14}, \citealt{Sakari_15}).
While Terzan~7 and Palomar~12 are commonly associated with the currently disrupting Sagittarius dwarf galaxy (see \citealt{Law_10} and references therein), the other clusters are orphans without a known parental system. Only because of their abnormal chemical signatures is it speculated that they have formed in smaller satellite systems with slower chemical enrichment, and have subsequently been stripped from their hosts during the accretion (e.g., \citealt{Villanova_13}, \citealt{Colucci_14}).
In this scenario, the GCs \emph{need} to share the fingerprint of slow chemical enrichment inherent to the host galaxy. Furthermore, it would be possible to generally use clusters with peculiar alpha-abundances for chemical tagging of accreted systems.

The only Local Group galaxy known to host GCs while its stars are simultaneously alpha-depleted at a metallicity of Ruprecht~106 or G002 is the Fornax dwarf spheroidal. This galaxy therefore may provide a unique test case to address the key questions of whether GCs share the chemical enrichment pattern of their host galaxy and, as a consequence, whether it is possible to use peculiar alpha-abundances (i.e., those deviating from the field stars) for chemical tagging of accreted systems.

\begin{figure}[h,t]
\begin{center}
\includegraphics[width=0.5\textwidth]{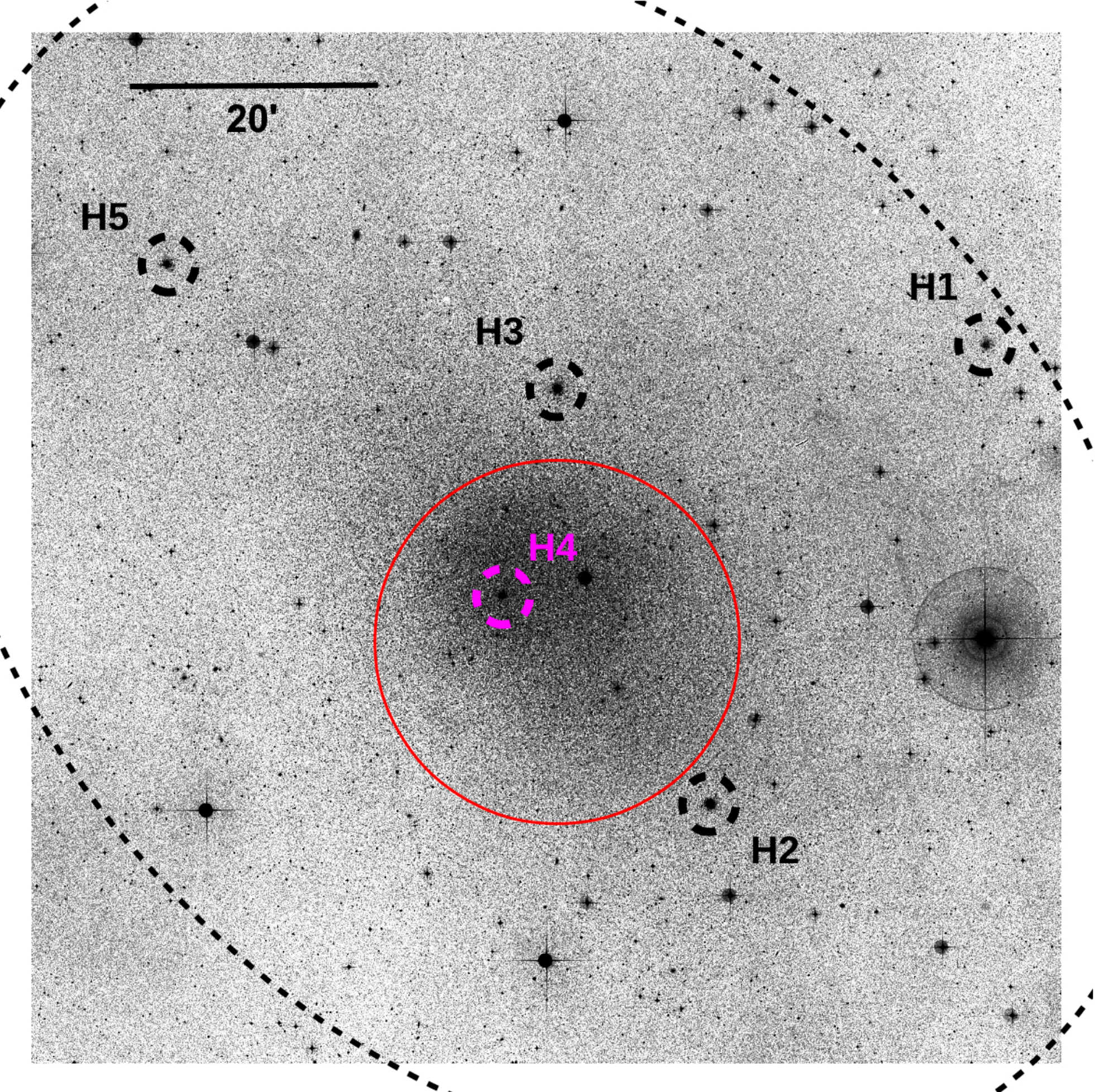}
\caption{DSS image of the Fornax dwarf spheroidal galaxy. The position of the five GCs are highlighted. H4 is located close to the center of the galaxy and therefore suffers from strong contamination of the surrounding field. The red circle shows the approximate pointing and field-of-view of \textit{M2FS} with a radius of $14.65 \arcmin$. The dashed ellipse indicates the tidal radius of the galaxy.}
\label{fig_2_1}
\end{center}
\end{figure}

Fornax is considered a ``classical'' dwarf spheroidal galaxy. It is one of the most massive Galactic satellites with a few $10^7\,M_{\sun}$ (\citealt{McConnachie_12}), an extended star formation history (\citealt{de_Boer_12}, \citealt{del_Pino_13}), and consequently hosts stars of a broad range of metallicities (\citealt{Battaglia_06}, \citealt{Hendricks_14b}). Moreover, the galaxy hosts its own population of five GCs (see Figure~\ref{fig_2_1}). Four of the Fornax GCs are metal-poor with [Fe/H]$\leq -2.0$ (\citealt{Letarte_10}), old (\citealt{Buonanno_98}) and alpha-enhanced (\citealt{Letarte_10}, \citealt{Larsen_12}), and by that resemble typical MW halo clusters. 
The remaining cluster (named H4, following \citealt{Hodge_61}) is an outlier in many respects: it is significantly more metal-rich, around [Fe/H]$=-1.4$ (\citealt{Strader_03}, \citealt{Larsen_12}), and possibly younger than the other clusters (\citealt{Buonanno_99}). 
Most importantly, it has only recently been shown that field stars in Fornax become alpha-depleted at very low metallicities with a knee in the alpha-evolution most likely below [Fe/H]$\approx-1.9$\,dex (\citealt{Hendricks_14a}, \citealt{Lemasle_14}). As a consequence, H4 is located at a metallicity where the field stars in Fornax are clearly alpha-depleted. This contrasts with the composition characteristics of similar metallicity MW field stars, which are alpha-enhanced. Therefore, H4 provides a unique opportunity to test if GCs share the chemical enrichment pattern of their host galaxy, and if it is possible to use peculiar alpha-abundances to chemically identify accreted systems such as Ruprecht~106 in the MW or G002 in M31.

Despite its importance for understanding GC formation and evolution in Fornax, H4 is the only cluster in the galaxy for which no individual member stars have been analyzed to date. H4 is located very close to the center of the galaxy and is therefore severely contaminated by field stars. Additionally, it happens to be the most compact of all Fornax GCs, and at a distance of $\sim147$\,kpc its core radius amounts to only $2.64\arcsec$ (\citealt{Mackey_03}). From ground-based telescopes, H4 is only resolved into individual stars in its outer regions where 
the fraction of cluster member and field star contaminants is about equal.

In the study presented here, three likely individual member stars of the cluster H4 have been identified. For one of them, we derived detailed chemical abundances and put them in direct comparison with field stars of both the Fornax dwarf spheroidal and the MW at similar metallicity. The goal of this work is to shed light on the question of whether or not the chemical enrichment history of field stars within a galaxy is imprinted on its GC system.

The paper is organized as follows.
In Section~\ref{chap_02}, we summarize our target selection, observing setup, and data reduction. Section~\ref{chap_03} gives details about the chemical analysis, which is subsequently presented in Section~\ref{chap_04} together with an age analysis of H4. In Section~\ref{chap_05}, we discuss the impact of our findings with respect to the nature of alpha-depleted GCs, the nature of H4 itself, and the chemical enrichment properties of Fornax. Finally, in Section~\ref{chap_06}, we summarize our main results.

\section{Data acquisition and reduction} 
\label{chap_02}

Finding individual and bright member stars in H4 is a challenging endeavor. A multi-object spectrograph is necessary in order to efficiently observe 
cluster members. Furthermore, the cluster's large distance with accessible targets located at $10 - 60\arcsec$ from the cluster center, means that the
instrument must be capable of densely packing many objects onto a small spatial scale. To avoid contamination, the apertures also need to be small and able to be placed on the field-of-view with high precision. Finally, with most targets fainter than $V=19.0$\,mag, exposure times to obtain sufficient signal-to-noise for a chemical analysis are long and suffer from accumulating cosmic rays.

\subsection{Observations, instrument, and setup}

For this project, we used the Michigan/Magellan Fiber System (\textit{M2FS}, \citealt{Mateo_12}) and MSpec spectrograph, a fiber-fed spectrograph mounted on the Nasmyth-east port of the Magellan-Clay 6.5m telescope at Las Campanas Observatory. 
For \textit{M2FS}, we used the $Bulge\_GC1$ setup and 125$\,\mathrm{\mu}$m slit (see \citealt{Johnson_15}), which allows a simultaneous observation of up to 48 targets at a resolving power of $R\approx 28\,000$ and a continuous wavelength coverage from $6140$ to $6720$\,\AA. 
The observations have been carried out during three consecutive nights in December 2014. With 8 individual exposures, a total of 6.7\,hours have been observed on target with typical seeing conditions around $0.8\arcsec$.

\subsection{Target selection}

Due to the central location of H4 in the field of the Fornax dwarf galaxy, the cluster is heavily contaminated by Fornax field stars. The contamination fraction in the outer half of the cluster tidal radius is higher than 50\%, and still amounts to $\geq20$\% around the cluster center (see Section~\ref{chapter_membership_selection}). A careful target selection is critical to maximize the fraction of actual H4 members, while simultaneously avoiding blended stars in the heavily crowded area. 
In addition to bona-fide H4 cluster members, we also deploy some of the fibers on bright field stars in the vicinity of the cluster to allow a direct comparison between cluster and galactic properties.

\subsubsection{GC candidate member stars}
\label{chapter_target_selection_1}

For our target selection, we used archival Hubble Space Telescope imaging taken with the Wide-Field Planetary Camera 2 (WFPC2) in programme 5637 (PI: Westphal). 
The data consist of two deep (1100\,s) and one shallow (60\,s) exposure in both of the F555W (V) and F814W (I) bands, and were first published 
by \citet{Buonanno_99}. We retrieved the pipeline-reduced individual images from the STSCI archive and performed point-spread 
function fitting photometry using \textsc{hstphot} \citep[version 1.1; ][]{Dolphin_00}, following the same procedure described in more 
detail in \citet{Frank_12}. Briefly, residual shifts of the images were determined by initially running \textsc{hstphot} on each 
image separately and cross-matching the resulting catalogs. After the creation of cosmic-ray masks, the well-aligned deep exposures in each 
filter were co-added, and the F555W deep co-added image was used as a detection image in the simultaneous photometry from all frames. 

The output of \textsc{hstphot} provides charge-transfer-efficiency- and aperture-corrected magnitudes in the HST system (F555W and F814W), as 
well as magnitudes transformed to the Johnson-Cousins V and I bands, based on the updated calibration and photometric transformation 
coefficients of \citet{Dolphin_09}. In the following, we use these V and I band magnitudes.

\begin{figure}[h]
\begin{center}
\includegraphics[width=0.5\textwidth]{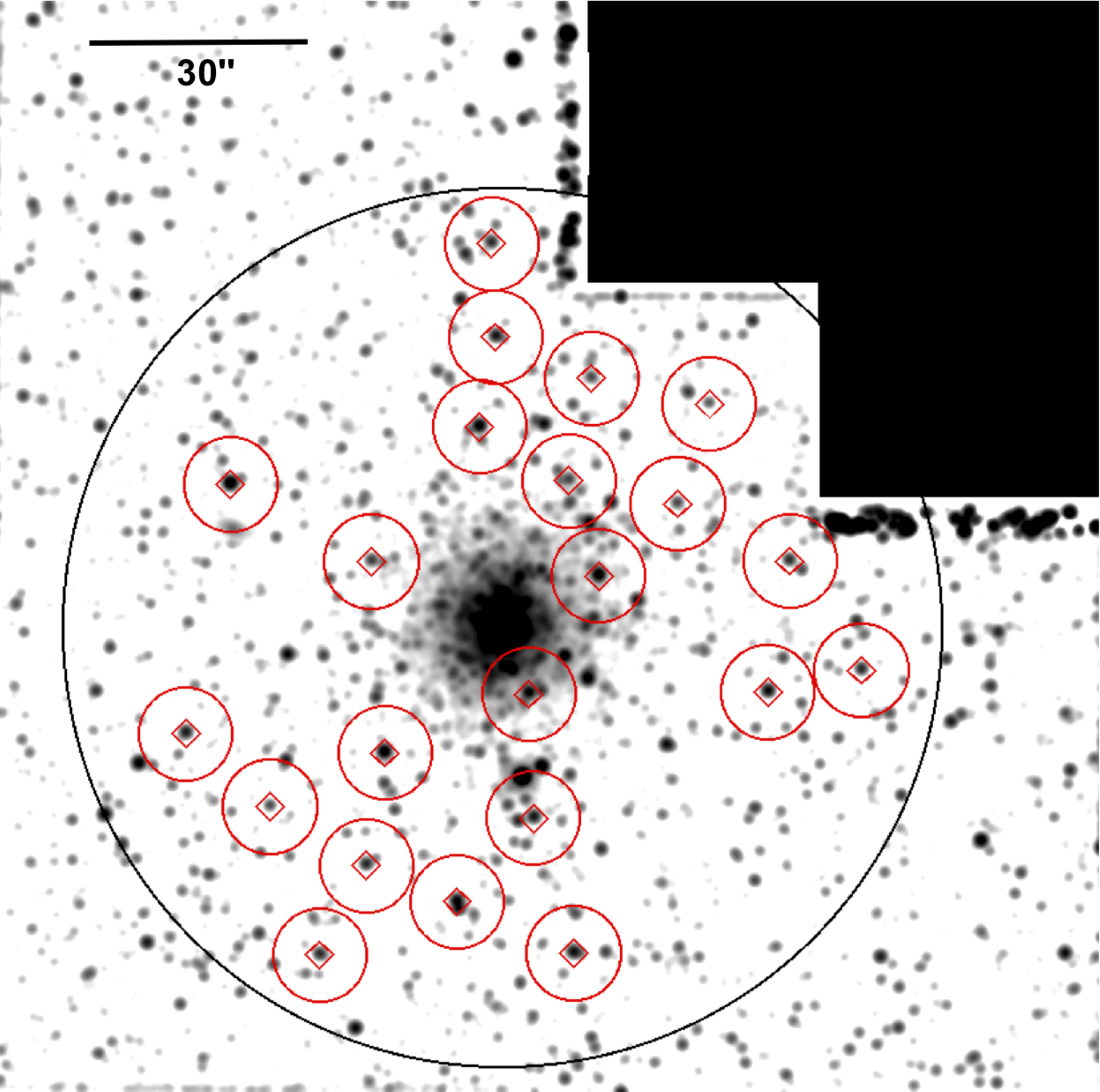}
\caption{Location of our targets in the GC H4, shown on the WFPC2 F555W-band mosaic and artificially degraded to a ground-based seeing of $1\arcsec$. The tidal radius of the cluster ($r_{t}=1\arcmin$) is shown in black and our targets are highlighted with red symbols. Red boxes are $2\arcsec \times2 \arcsec$ and mimic the actual fibersize while the circles are $6.5\arcsec$ in radius and visualize the minimum allowed spacing between individual fibers for \textit{M2FS}.}
\label{fig_2_2}
\end{center}
\end{figure}

For astrometry, we created a mosaic of all F555W-band exposures (see Figure~\ref{fig_2_2}) using \textsc{multidrizzle} 
\citep{Koekmoer_06}, and transformed the photometric catalogs of the camera's four individual chips to this reference image, in order to 
correct for geometric distortion. We corrected for shifts in absolute R.A. and Dec by matching our catalog to the 2MASS astrometric system 
\citep{Cutri_03}, using the VIKING survey DR1 source catalog (\citealt{Edge_13}) as an intermediate step in the cross-match
because the 2MASS point source catalog is very sparse in the $\sim 2.6\arcmin \times 2.6\arcmin$ field-of-view of the WFPC2 pointing.

Possible targets are selected from the red giant branch (RGB) and comprise stars from the RGB-tip down to magnitudes as faint as $V=20.5$\,mag. Although the RGB is fairly broad with a possible color split for the brightest stars, it is not clear whether this is a signature of the GC population on top of a field stars mix or whether it simply reflects the intrinsic spread in the field star population.
Therefore, we did not further constrain our selection to a specific part of the RGB, as for example done in \citet{Carretta_10} for the case of M54 in the center of the Sagittarius dwarf spheroidal, where a cleaner separation between populations could be made.
We also refrain from placing an individual fiber at the (unresolved) center of the cluster. The analysis of the resulting integrated-light spectrum would be inferior to classical drift-scan methods because the integrated spectrum does not consist of an entire population (which can be modelled at varying levels of accuracy) but may contain only some dozens of stars of essentially unknown origin and parameters. 

\begin{figure}[b]
\begin{center}
\includegraphics[width=0.5\textwidth]{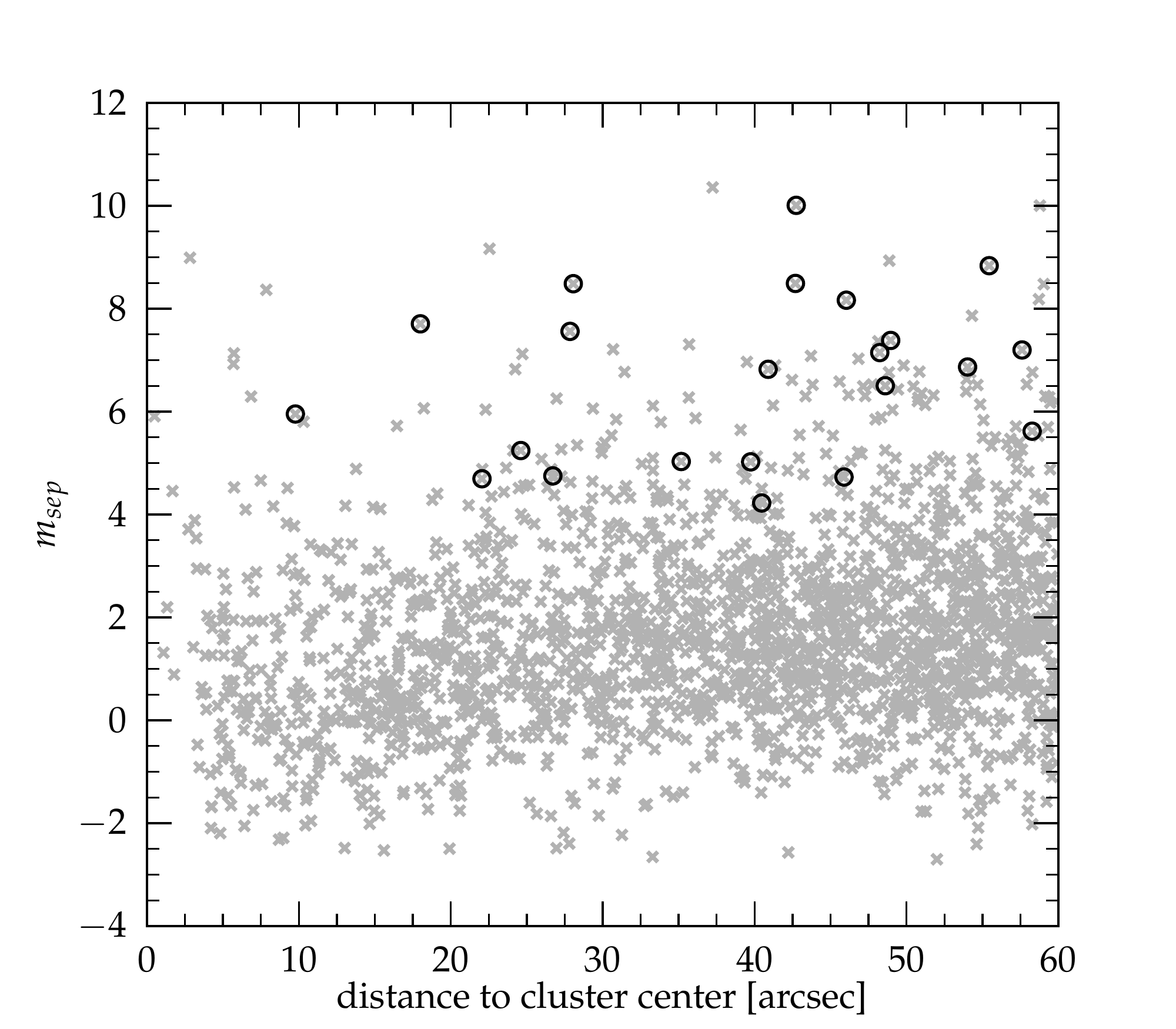}
\caption{Plot of the separation index defined in the text for stars in the GC H4 as a function of central distance. Gray crosses denote all red giant stars projected on the sky near the cluster, and our targets are highlighted with black circles. While $m_{sep}=0.0$ means that half of the star light comes from blending neighbors, $m_{sep}=4.0$ and $5.0$ means $\sim2.5$\% and $1.0$\% contamination, respectively.}
\label{fig_2_3}
\end{center}
\end{figure}

Blending is a serious problem for stars in H4, specifically in the inner regions of the cluster where membership likelihood is highest, and we have to take into account the additional seeing from ground-based observatories in contrast to the HST images. To quantify the amount of blending and to exclude significantly flux-contaminated stars in advance, we calculate a \textit{separation index}, developed by \citet{Stetson_03}.
In detail, we use the magnitude of each star in the HST catalog to calculate its flux, which we subsequently smooth with a Gaussian profile according to the seeing of our observations.
We then calculate the flux ratio between the target star and all neighbors in the environment at the star's central position and express the result as a magnitude ($m_{sep}$).
Finally, we exclude all stars from our target list with a flux contamination higher than 5\%, equivalent to $m_{sep} \geq 3.25$. Our final targets typically have $m_{sep} \geq 5.0$, and by that suffer less than 1\% contamination from neighboring stars (see Figure~\ref{fig_2_3}).

\subsubsection{Field stars}

Field star targets of the Fornax dwarf spheroidal were selected from the catalog of \citet{Battaglia_08}, which provides Calcium Triplet metallicities for nearly $1\,000$ bright RGB stars that are cleaned from most of the contaminating Galactic halo foreground by velocity cuts.
From this catalog, we purposefully picked rare, metal-poor stars with [Fe/H]$\leq-1.4$\,dex to trace the alpha-element evolution at the metallicity of H4 and at the expected position of the [$\mathrm{\alpha}$/Fe] ``knee''.
The selected field stars belong to the upper part of the RGB and consequently have similar magnitudes to the H4 candidate stars. 
Since they are located close to the cluster itself, our field star targets are all located in the central part of Fornax ($r\leq0.3\degree$) and hence within the core radius of the galaxy. 

\subsection{Data reduction}
\label{chapter_datareduction}

The general spectroscopic data reduction process follows the description in \citet{Johnson_15}. 
Using IRAF routines, the individual amplifier images on each CCD were separately trimmed, bias corrected and subsequently rotated, translated, and combined into one exposure per CCD.
Next, we used the IRAF task \textsc{dohydra} to extract the individual orders of each spectrum.
This task has been written to extract spectra taken with the WIYN and Blanco Hydra spectrographs, and it can be applied to most multifiber echelle datasets. The routine includes aperture identification and tracing, scattered light removal, flat-field correction, throughput- and wavelength-calibration, and a basic cosmic-ray removal. 
Sky-subtraction within \textsc{dohydra} is skipped and performed later using a master-sky frame from the combined sky fibers of both CCDs.

It is important to emphasize that \textsc{dohydra} is called separately on each of the six orders for individual spectra. As a consequence, the extracted parts of any full spectrum underwent an individual data reduction process and therefore possibly display different systematic signatures imprinted by the individual extraction steps. This allows us to verify the 
robustness of our analysis against such possible systematics by comparing the results from different orders (see Section~\ref{consistency_tests}).

After sky subtraction, the individual exposures have been summed using a weighted average based on the typical S/N of the brightest targets in each frame, to maximize combined signal-to-noise. 
The heliocentric velocity was removed from each frame prior to co-addition in order to account for differences of up to $0.7$\,$\mathrm{km\,s^{\rm -1}}$ between individual frames.
Finally, the spectra are continuum normalized and the full wavelength range is recovered by combining the individual orders.

Compared to the data in \citet{Johnson_15}, our targets are extremely faint with long individual exposure times and low signal-to-noise ($\leq10$ per pixel) within the individual exposures. This introduces some complications in the data reduction and requires some additional steps.
First, given the long individual exposure times of up to 1~hour, the images suffer from severe cosmic-ray contamination for which \textsc{dohydra} cannot properly account. The multitude of cosmic-ray features not only possibly spoil absorption features in the spectra, but also may pose a problem for fiber tracing, throughput calibration, and later to sky subtraction and continuum placement.
For this reason, we use the Python implementation of \textsc{LACosmic}\footnote{available online at {http://obswww.unige.ch/$\sim$tewes/cosmics\_dot\_py}} (\citealt{vanDokkum_01}) on our 2d images prior to the \textsc{dohydra} task, which yields a significant improvement on our results.
However, we cannot be sure if absorption features, initially affected by cosmic-rays, are recovered to their intrinsic, unbiased properties. Therefore, we generate a cosmic-ray mask in order to flag any regions in a spectrum which have been possibly biased by cosmic-ray removal.
Unfortunately, \textsc{dohydra} does not propagate bad pixel masks through the reduction. We therefore perform the extraction process twice, once on the images with and once on the images without previous correction by \textsc{LACosmic}. 
Then, we flag all wavelength regions where we observe a significant ($\geq 5$ sigma of continuum noise) difference between the extracted versions.
Finally, we combine the regions from all individual exposures to build the cosmic-ray mask. Later, we will use this mask to test the robustness of our abundances by excluding these regions during the analysis (see Section~\ref{consistency_tests}). 

Despite the initial throughput calibration, we observe flux variations between individual fibers, including science and sky fibers.
This necessitates a sophisticated sky-subtraction for which we use the \textsc{skytweak} task in IRAF. The \textsc{skytweak} task allows for wavelength shifts and rescaling of the flux in the master-sky spectrum to minimize residuals of the most prominent emission features. 
The master-sky is generated by the average of 5 sky fibers distributed among both CCDs, and includes a min-max rejection algorithm to clear the spectrum of any remaining cosmic-rays or other contamination.
In low signal-to-noise spectra with flux only $\sim10$ times above the sky level, sky residuals will have a non-negligible impact on any chemical analysis. 
Similar to the cosmic-ray mask, we therefore generate a \textit{sky-mask} where we flag wavelength regions with initially strong sky emission lines and test the sensitivity of our chemical analysis to these regions.
The reddest order in our setup (order\,57; $6653$ - $6720$\,\AA) does not contain any significant sky emission features that can be used for rescaling. As a consequence, the level of sky subtraction is less reliable than for the other orders, and we therefore exclude this region from the chemical analysis for all stars.

Finally, some orders of some spectra are affected by an internal Littrow ghost reflection (e.g., \citealt{Burgh_07}; see also \citealt{Johnson_15}, their Figure~1). Such regions are clearly visible as a strongly deviating continuum flux, and we exclude these regions from the forthcoming analysis.

After data reduction, we obtain a total of 40 science spectra (18 field stars and 22 stars within the tidal radius of H4). The signal-to-noise distribution, measured at $\sim 6388$\,\AA, ranges from $\sim40$ per pixel for the brightest targets to $\leq10$ for the faintest objects. We estimate the final resolution of our spectra from the width of several clean sky emission lines and find a constant $R={{\lambda} / {\Delta \lambda}} \approx 28\,000$ over the full wavelength range.

\section{Data analysis}
\label{chap_03}

\subsection{Radial velocities}

We determine radial velocities for each star using he IRAF task \textsc{fxcor} and a template spectrum convolved to the resolution of our observations.
As template we used a synthetic spectrum\footnote{Obtained from the \citet{Coelho_05} library of high-resolution synthetic stellar spectra.} with $\mathrm{T_{eff}}=4250$\,K, $\log g=1.0$, and [Fe/H]$=-1.4$, which is close to what we expect for H4 cluster members and suitable for our complete sample. 

To obtain the most precise velocity determination, we use only one of the orders (order\,55; 6406 - 6515\AA) because it contains a series of deep, unblended absorption lines, it is relatively free of strong sky emission, and it does not contain broad lines nor strong telluric bands.
Where applicable, we test the outcome with other orders and find consistent results within the uncertainties.

Velocity uncertainties reported by \textsc{fxcor} are typically $\leq 1.0$\,$\mathrm{km\,s^{\rm -1}}$ (between 0.2 and 2\,$\mathrm{km\,s^{\rm -1}}$, depending on S/N). We additionally estimate the uncertainty from the standard deviation of the individual exposures for each star, which yield very similar results to the \textsc{fxcor} values.
Although less critical than for the chemical analysis, we want to make sure that our velocities are not strongly biased by residual artifacts from sky emission or cosmic rays. Therefore, we determine velocities once using the full wavelength range of order 55 and once without the masked regions in the sky- and cosmic-ray mask. Reassuringly, the differences are small and usually at or below the level of the statistical errors.

\subsection{Chemical analysis}
\label{chap_chemical_analysis}

\subsubsection{\textsc{SP\_Ace}}

For the chemical analysis of our spectra we use the newly developed code \textsc{SP\_Ace} (Stellar Parameters and Chemical Abundances Estimator; \citet{Boeche_15}), an evolution of the RAVE chemical abundance pipeline (\citealt{Boeche_11}). In short, the code employs a full-spectrum-fitting technique to derive stellar atmospheric parameters and chemical abundances by means of a $\chi^2$-minimization procedure. The reference spectra are generated with a library of general curve-of-growth functions for a given line-list within the desired wavelength range. We point the interested reader to \citet{Boeche_15} for a detailed description of the method, the code itself, and its performance on synthetic and real spectra.

A major advantage of fitting entire wavelength regions is that the amount of information extracted from the data is improved compared to an individual line analysis, specifically because blended absorption features of the same and of different species can be incorporated in the analysis. 
As a consequence, reliable results can be obtained for many chemical elements, even in spectra of low resolution and/or low signal-to-noise (e.g. \citealt{Kirby_08}, \citealt{Caffau_13}, \citealt{Conroy_14}, \citealt{Choi_14}, or \citealt{Hendricks_14a}).
Another advantage is that the iterative fitting procedure of the full spectrum within \textsc{SP\_Ace} is able to take into account the knowledge about individual absorption features for the continuum placement, which otherwise can have an unwanted biasing effect on the derived continuum level and therefore on the derived chemical abundances (e.g. \citealt{Kirby_08}).
The limited luminosity (and hence distance) range in which classical, high-resolution, and high signal-to-noise chemical abundance analysis can be carried out can thus be expanded to extragalactic targets, such as MW satellites or unresolved systems in the Local Group. 

While it seems generally possible to perform a chemical analysis with $\sigma\mathrm{[X/Fe]}\leq0.2$ even on low-resolution spectra with $R\sim3000$ (e.g. \citealt{Kirby_08}, \citealt{Conroy_14}), \textsc{SP\_Ace} is theoretically capable to determine robust abundances for $R\geq2000$ and is specifically tested between $R=2000 - 20\,000$. Therefore, we degrade our spectra slightly to a resolution of $R\approx16\,000$ to place them on the well-calibrated regime of \textsc{SP\_Ace} and also to obtain a higher signal-to-noise per resolution element. For the fainter stars, this improves the proper placement of the continuum and reduces the confusion between noise and weak absorption lines within the fitting procedure.
Nonetheless, we test the consistency of the derived chemical abundances to results obtained using the original $R\approx28\,000$ resolution, and find low scatter and no systematic changes in the results (see Section~\ref{consistency_tests}).

The most critical problem for an automated, full-spectrum analysis routine is its susceptibility to artifacts related to fitting sky and cosmic ray residual features, which may lead to problems in the continuum placement. Spectra of lower signal-to-noise, as we analyze here, are specifically vulnerable to these points.
To address this issue, we visually inspect the model fit to each spectrum and remove individual pixels or wavelength areas from the fit in cases where they can be clearly identified as the origin for a mismatch.
This ``visual mask" is iteratively refined for each of our targets until we either obtain a satisfactory fit or no obvious cause for a poor fit can be identified. In the latter case, the spectra will have large $\chi^2$-values, and we treat their outcome with care.
On top of this, we reject absorption features that are problematic to model (in our wavelength range, e.g., $H_{\alpha}$ or the Telluric feature blueward of 6320\,\AA).

In summary, we supply four different pixel masks to \textsc{SP\_Ace} which define wavelength regions to ignore in the analysis. 
While use of the visual mask and the regions containing the Littrow ghost feature are mandatory in order to obtain reliable results, the cosmic-ray and sky 
masks only have a negligible impact on our results. We only use the cosmic-ray and sky masks later to test the robustness of the chemical analysis against possible inaccuracies in the correction for sky emission and cosmic ray contamination during the process of data reduction.

\subsubsection{Atmospheric parameters}
Generally, \textsc{SP\_Ace} is capable to derive stellar atmospheric parameters within the fitting process. However, given the fairly low signal-to-noise of our spectra, a more robust result can be achieved when $\mathrm{T_{eff}}$ and $\log g$ are estimated from multi-band photometry.
Specifically, we use $V$ and $I$ magnitudes from our HST photometry and derive temperatures $\mathrm{T_{eff}}$ and bolometric corrections from the empirical equations given in \citet{Alonso_99}.
For the reddening, we adopt $E(B-V)=0.08$ from \citet{Greco_07} and apply an object-specific transformation described in \citet{Hendricks_12} (specifically their Eq.~3 and 4) to obtain $E(V-I)=0.11$, suited for a red giant of [Fe/H]$=-1.0$\,dex, $\mathrm{T_{eff}}=4250$\,K, and $\log g=1.5$. Adopting a standard reddening law ($R_{V}=3.1$), we furthermore obtain a $V$-band extinction of $A_{V}=0.248$. We assume no significant differential reddening within our field-of-view.

For surface gravities, $\log g$, we then used the standard relation:

\begin{equation}
 \log g = \log g_\odot + \log \left[\left(\frac{T_\mathrm{eff}}{T_\mathrm{eff,\odot}}\right)^4 \left(\frac{M}{M_\odot}\right) \left(\frac{L_\mathrm{bol}}{L_\mathrm{bol,\odot}}\right)^{-1}\right] .
 \label{eq:logg}
\end{equation}

Here, we apply a distance modulus of $\mu_{0}=20.84$, adopted from \citet{Pietrzynski_09}. In order to account for a possibly large age spread among stars in our sample, we used Dartmouth Isochrones (\citealt{Dotter_08}) and a simple linear age-metallicity relation to assign individual masses to our targets. Specifically, we obtain a variation of stellar mass from $1.27$\,$M_{\sun}$ for stars with [Fe/H]$=-1.0$ to $0.79$\,$M_{\sun}$ for a stars with [Fe/H]$=-2.0$. Finally, the micro-turbulence is assumed to be a function of $\mathrm{T_{eff}}$ and $\log g$, and is calculated with a third-order polynomial given in \citet{Boeche_15}. 
The atmospheric parameters are iteratively determined with updated stellar metallicities from \textsc{SP\_Ace}.

\subsubsection{Consistency tests}
{\label{consistency_tests}}

As outlined in the previous sections, several of the assumptions we make during data reduction and analysis may have systematic effects on the derived chemical abundances. Therefore, we perform a series of consistency tests to assess the robustness of our analysis against these factors. Specifically, we test:

\begin{compactenum}[a)]

\item the results obtained using the full wavelength range (minus the mandatory pixel masks) compared to the outcome using only one of the five individual orders. The motivation of this test is that each order underwent an individual reduction and extraction process, including wavelength calibration, throughput calibration, (tweaked) sky subtraction, and continuum fit, and therefore may display individual systematic biases.

\item the results obtained using the full wavelength range compared to the outcome when additional sky emission regions are flagged out, using the sky-mask. The test is motivated by the necessity for a tweaked sky subtraction to compensate for inaccurate throughput calibration.

\item the results obtained using the full wavelength range compared to the outcome when additional cosmic ray regions are flagged out, using the cosmic-ray mask. The test is motivated by the necessity for a cosmic-ray subtraction prior to spectrum extraction, due to the long individual exposure times.

\item the results obtained using a slightly degraded resolution of $R\approx16\,000$, compared to the original $R\approx28\,000$.

\item the results obtained using artificially altered atmospheric parameters by $\Delta \mathrm{T_{eff}}=+200$\,K.

\item the results obtained using artificially altered atmospheric parameters by $\Delta \log\,g=+0.3$.

\end{compactenum}

\begin{figure*}[t]
\begin{center}
\includegraphics[width=1.0\textwidth]{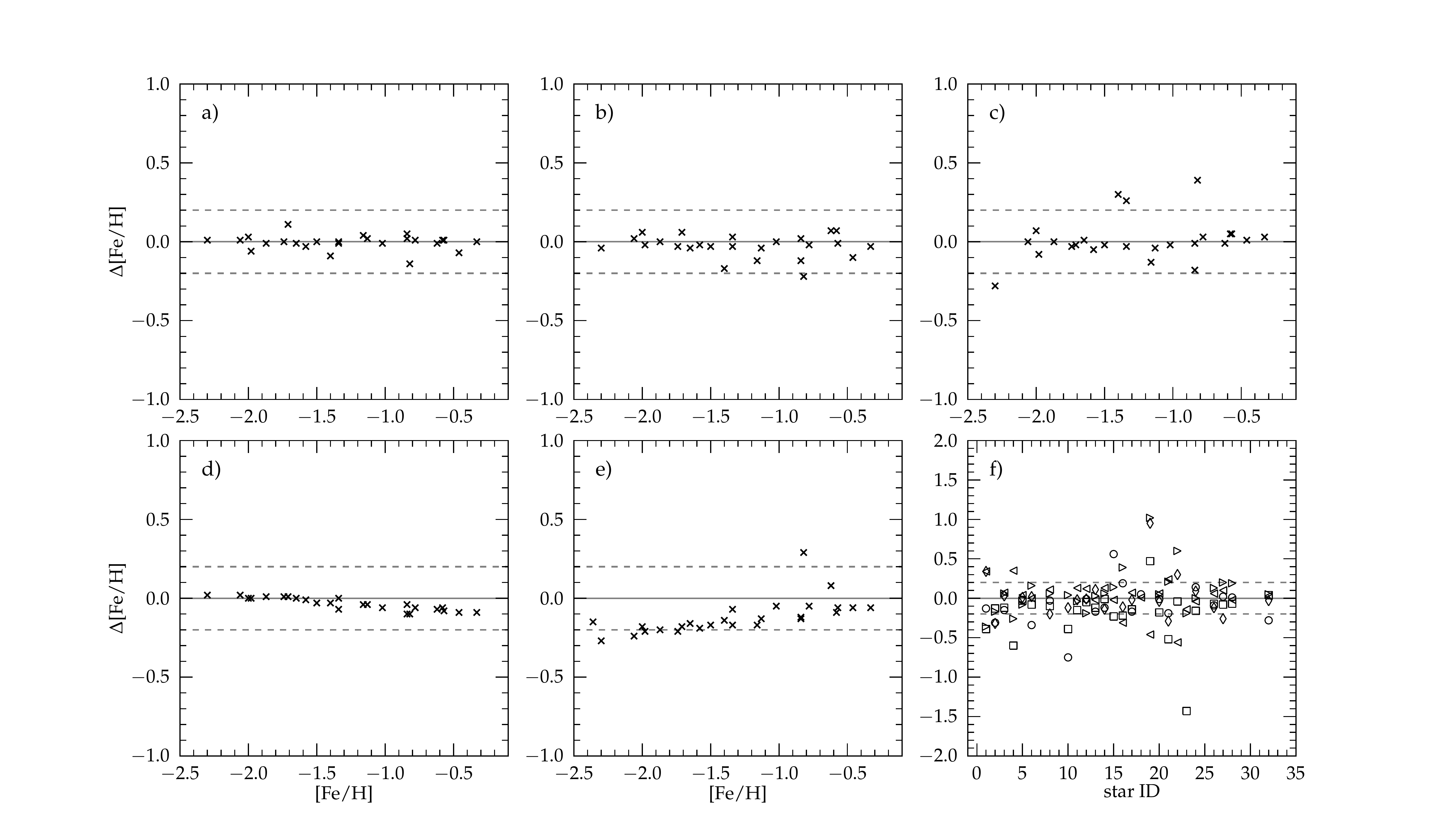}
\caption{Consistency tests for chemical abundances derived with \textsc{SP\_Ace}. All plots show the difference between the derived [Fe/H] and various changes to the spectrum as described in the text:
\textbf{a)} with cosmic ray regions flagged out;
\textbf{b)} with sky emission regions flagged out;
\textbf{c)} adopting the initial resolution of $R=28\,000$;
\textbf{d)} $\Delta \log\,g=0.3$;
\textbf{e)} $\Delta \mathrm{T_{eff}}=200\,K$;
\textbf{f)} all five orders analyzed separately (orders 52, 53, 54, 55, 56: circle, diamond, square, left-handed triangle, right-handed triangle). Given the limited wavelength range and the limited chemical information, not every individual order in each spectrum provides a result. Additionally, some orders in some stars are more affected by the different pixel masks and hence might show larger offsets.}
\label{fig_3_1}
\end{center}
\end{figure*}

Results for [Fe/H] are shown in Figure~\ref{fig_3_1}. We did not 
find any global systematic bias in the derived abundance for any of the 
tested scenarios (except for the artificial temperature and gravity variations).
We also did not find any trends with metallicity, and the scatter for all cases is typically well within $\pm0.2$\,dex. Importantly, all five individual orders yield consistent results, but with some scatter caused mostly by the limited line information within the smaller wavelength range.
A change in $\mathrm{T_{eff}}$ by $+200$\,K results in $\Delta$[Fe/H]$\approx-0.2$\,dex for stars with low metallicity and decreases to smaller offsets towards more metal-rich stars. A change in $\log~g$ by $+0.3$ has practically no effect amongst metal-deficient stars, and yields $\Delta$[Fe/H]$\approx-0.1$ for stars of high metallicity.

\subsubsection{Uncertainty estimates}

Statistical uncertainties are estimated within \textsc{SP\_Ace} and expressed as a separate upper and lower limit of the derived [X/H] elemental abundances ([X/H]$_{+}$ and [X/H]$_{-}$)\footnote{When the upper or lower abundance limit falls beyond the internal \textsc{SP\_Ace} abundance grid, a null-value is reported.}.
Here, the upper and lower limit express the 68\% probability with no guarantee that the semi-error expresses the 34\% limit (see \citet{Boeche_15} for more details). For the errors in relative abundances ([X/Fe]), we simply calculate the semi-errors as quadratic sum of the upper and lower semi-errors (e.g., $ \sigma\mathrm{[X/Fe]_{+}}^2=\sigma\mathrm{[X/H]_{+}}^2 + \sigma\mathrm{[Fe/H]_{+}}^2$) and therefore overestimate the error slightly by ignoring the covariance terms between [Fe/H] and [X/H] in a full-spectrum-fitting approach.
Uncertainties for [Fe/H] range from $\leq0.1$\,dex to $\sim0.3$\, dex, depending on the brightness and the metallicity of the star. For [$\mathrm{\alpha}$/Fe] (where $\mathrm{\alpha} =$ Ca, Si, or Ti) we obtain similar but somewhat larger errors. Iron-peak element (V, Cr, Co, Ni) ratios typically show uncertainties as small as [Fe/H].
Finally, the statistical uncertainties are included with some systematic error as discussed in the previous section, which limits the final accuracy of our results to $\delta \mathrm{[X/H]}\sim 0.1$\,dex, even for the brightest targets.

Astrometric and photometric information, as well as the derived chemical abundances with their uncertainties for all our targets are given in the Appendix (Table~\ref{table_1}, ~\ref{table_2}, and~\ref{table_a_3}, respectively).

\section{Results} 
\label{chap_04}

\subsection{[Fe/H], radial velocities and membership likelihood}
\label{chapter_membership_selection}

Due to its position close to the center of Fornax, stars within the tidal radius of H4 are severely contaminated by field stars of the galaxy. A membership likelihood determination and a clear assignment of our program stars to either the cluster or the field is crucial because the chemical properties of H4 \emph{compared} to the field is a major goal of this study.

Here, we use three observed properties to determine the membership likelihood of a target stars to the cluster H4: The star's proximity to the cluster center ($r_{GC}$), its radial velocity $v$, and its iron abundance [Fe/H] (here $m$).
Generally, the probability that a given star with properties $\{P\}$ is a member of the cluster H4 is given by

\begin{equation}
P_{\mathrm{H4}}(\{P\}) = { {N_{*}(\{P\}\,|\,\mathrm{H4})} \over {N_{*}(\{P\}\,|\,\mathrm{H4}) + {N_{*}(\{P\}\,|\,\mathrm{field}) } } },
\end{equation} 

where ${N_{*}(\{P\}\,|\,\mathrm{H4})}$ is the number of H4 members with properties $\{P\}=p_1$, $p_2$, ... $p_i$ and similarly $N_{*}(\{P\}\,|\,\mathrm{field}) $ denotes the number of field stars which share the same parameter space.

\begin{figure}[h,t,b]
\begin{center}
\includegraphics[width=0.5\textwidth]{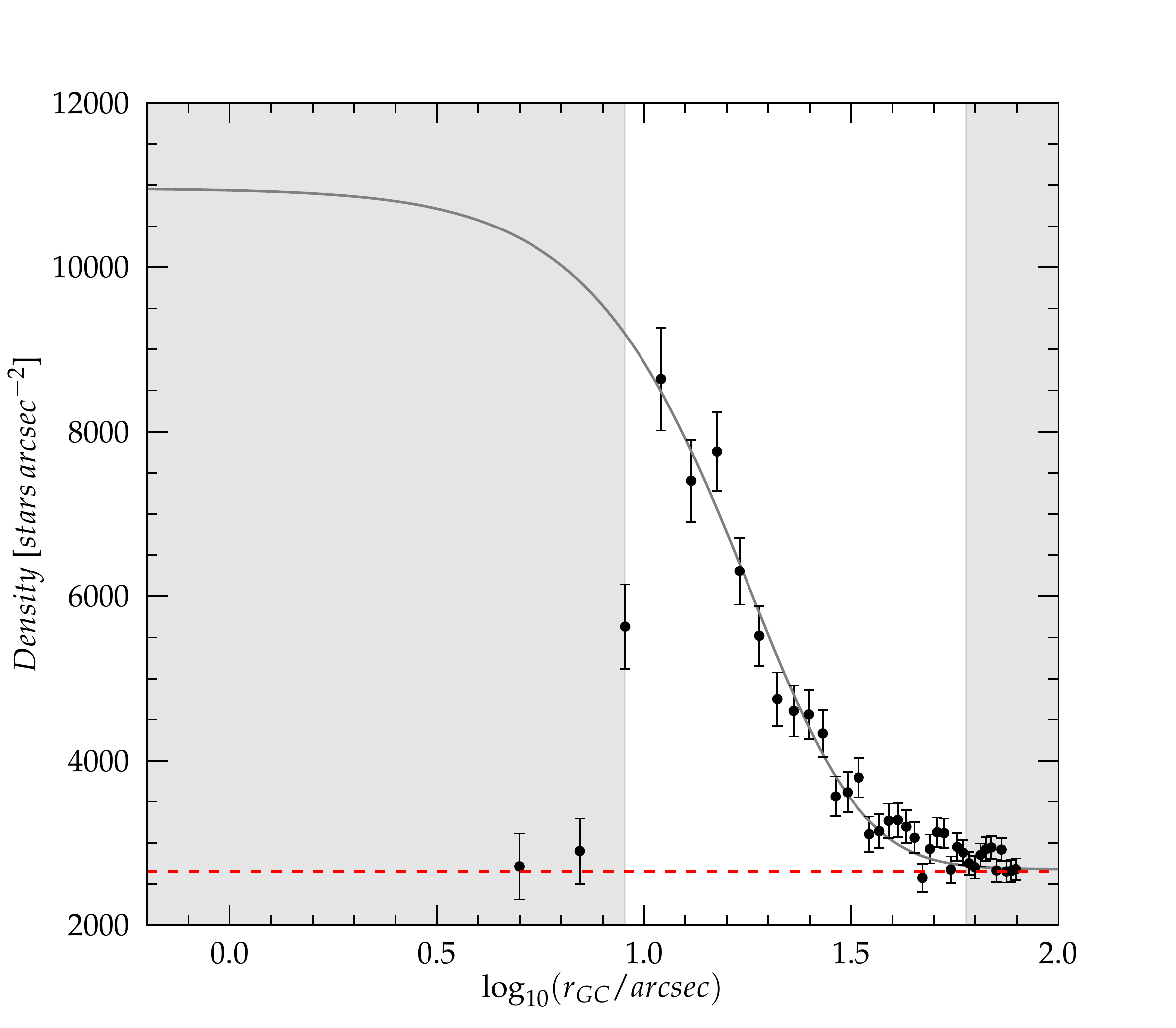}
\caption{Stellar density profile of the GC H4 from HST photometry. The gray line is a King-profile fit to the data at radii beyond $r_{GC}=12\arcsec$. The red line indicates the background level that we determine from the stellar density outside the tidal radius of the cluster. The white area in the figure shows the radial zone from which we pick our targets. At smaller radii, the cluster is not resolved with ground-based telescopes, and the probability that a star at larger radii is a member of the system is very low.}
\label{fig_4_1}
\end{center}
\end{figure}

Using our HST photometric catalog, we first obtain an initial membership probability from the stellar density profile of the cluster, which we fit with a King profile ($K(r_{GC})$; \citealt{King_66}), and extrapolate the function inwards to the unresolved center for radii $\leq12\arcsec$ (see Figure~\ref{fig_4_1}).
Notably, the contamination rate is high both at radii beyond half the tidal
radius (($\geq 50\%$) and near the cluster center ($\approx 20\%$).

The fraction of stars above the uniform background compared to the background level ($\rho_{b}$) itself yields the membership probability $P_{\mathrm{H4}}(r_{GC})$ for a given star with distance $r_{GC}$ to the cluster center:

\begin{equation}
P_{\mathrm{H4}}(r_{GC}) = { {N_{*}(r_{GC}\,|\,\mathrm{H4})} \over {N_{*}(r_{GC}\,|\,\mathrm{H4}) + {N_{*}(r_{GC}\,|\,\mathrm{field}) } } }= {{ K(r_{GC}) } \over { K(r_{GC}) + \rho_{b} } }.
\end{equation} 

The unfortunate circumstance of heavy contamination is somewhat compensated by the fact that H4 members show a distinctly different radial velocity and [Fe/H] compared to the field stars. The radial velocity of the cluster has been measured in integrated-light studies by \citet{Larsen_12} and \citet{Dubath_92}, who consistently report values of $46.2$ and $47.2$\,$\mathrm{km\,s^{\rm -1}}$, respectively. Therefore H4's radial velocity is lower by $\sim 9$\,$\mathrm{km\,s^{\rm -1}}$ compared to the mean galactic motion of Fornax, which approximately corresponds to the velocity dispersion of the galaxy at this metallicity (see \citealt{Hendricks_14b}).
Additionally, H4's metallicity is measured around [Fe/H]$=-1.4$ (\citealt{Larsen_12}, \citealt{Strader_03}), and is therefore significantly lower than the galaxy average of $\sim-0.9$\,dex in the central part of Fornax.

When we combine the metallicity $m$ , the velocity $v$, and the proximity to the cluster center $r_{GC}$, we can put a tight membership probability for our stars:

\begin{eqnarray}
P_{\mathrm{H4}}(r_{GC},v,m) &=& { {N_{*}(r_{GC},v,m\,|\,\mathrm{H4})} \over {N_{*}(r_{GC},v,m\,|\,\mathrm{H4}) + {N_{*}(r_{GC},v,m\,|\,\mathrm{field}) } } } \\ \nonumber
\\ \nonumber
&=& { { K(r_{GC}) \times p(v,m\,|\,\mathrm{H4}) } \over { K(r_{GC}) \times p(v,m\,|\,\mathrm{H4}) + \rho_{b} \times p(v,m\,|\,\mathrm{field}) } } \nonumber
 . 
\end{eqnarray} 

Here, $p(v,m\,|\,\mathrm{H4})$ denotes the probability for an H4 member star to display the properties $v$ and $m$. Similarly, and $p(v,m\,|\,\mathrm{field})$ is the equivalent expression for field stars.

We extract the chemical and dynamical properties for the contaminating field stars from the catalog provided in \citet{Battaglia_08}, which provides Calcium-Triplet metallicities and radial velocities for nearly $1000$ stars. To take into account radial variations of these properties within Fornax, we only consider $406$ objects which are located at similar radii to the cluster, and specifically stars with $r \leq 0.3\degree$.

We do not know the exact properties of H4 given that all information is derived from integrated-light analysis.
With the risk of systematic misplacement of its actual properties, we assume that all H4 members fall within certain limits of [Fe/H] and $v$ where $p(v,m\,|\,\mathrm{H4})=1$ , but do not have a preferred position within these intervals. Consequently, there is a zero likelihood to find a member outside these limits.
We set the allowed parameter space for H4 members between $-1.65 \leq \mathrm{[Fe/H]} \leq -1.3$ and $42.5 \leq v \leq 52.5$\,$\mathrm{km\,s^{\rm -1}}$, constrained by previous integrated-light measurements that take into account their measurement error and a possible systematic bias in the methodology. For the velocities, we additionally take into account H4's intrinsic velocity dispersion of $\sim 4.5$\,$\mathrm{km\,s^{\rm -1}}$ (\citealt{Larsen_12}, \citealt{Dubath_92}).

For the field stars, the distributions of [Fe/H] and the radial velocities are not Gaussian functions nor are they any other evident analytical shape. We therefore abandon the attempt to model the complex distribution in a combined parameter space. Instead, we assess the contamination fraction $p(v,m\,|\,\mathrm{field})$ empirically from the fraction of field stars that fall within the allowed parameter box of H4 and find $p(v,m\,|\,\mathrm{field})=0.020$.
Under these assumptions---and if its Fe and $v$ identifies a star as a potential cluster candidate---the probability for each of our targets to be an actual member of the cluster, becomes

\begin{equation}
P_{\mathrm{H4}}(r_{GC},v,m) = {K(r_{GC}) \over {\rho_{b} \times p(v,m\,|\,\mathrm{field}) + K(r_{GC})}}.
\end{equation} 

Finally, we use a bootstrapping technique to estimate the uncertainty for $P_{\mathrm{H4}}(r_{GC},v,m)$ by constructing a number of equally sized samples where each individual entry is altered randomly within its uncertainties in $v$ and $m$.

\begin{figure}[t,b]
\begin{center}
\includegraphics[width=0.5\textwidth]{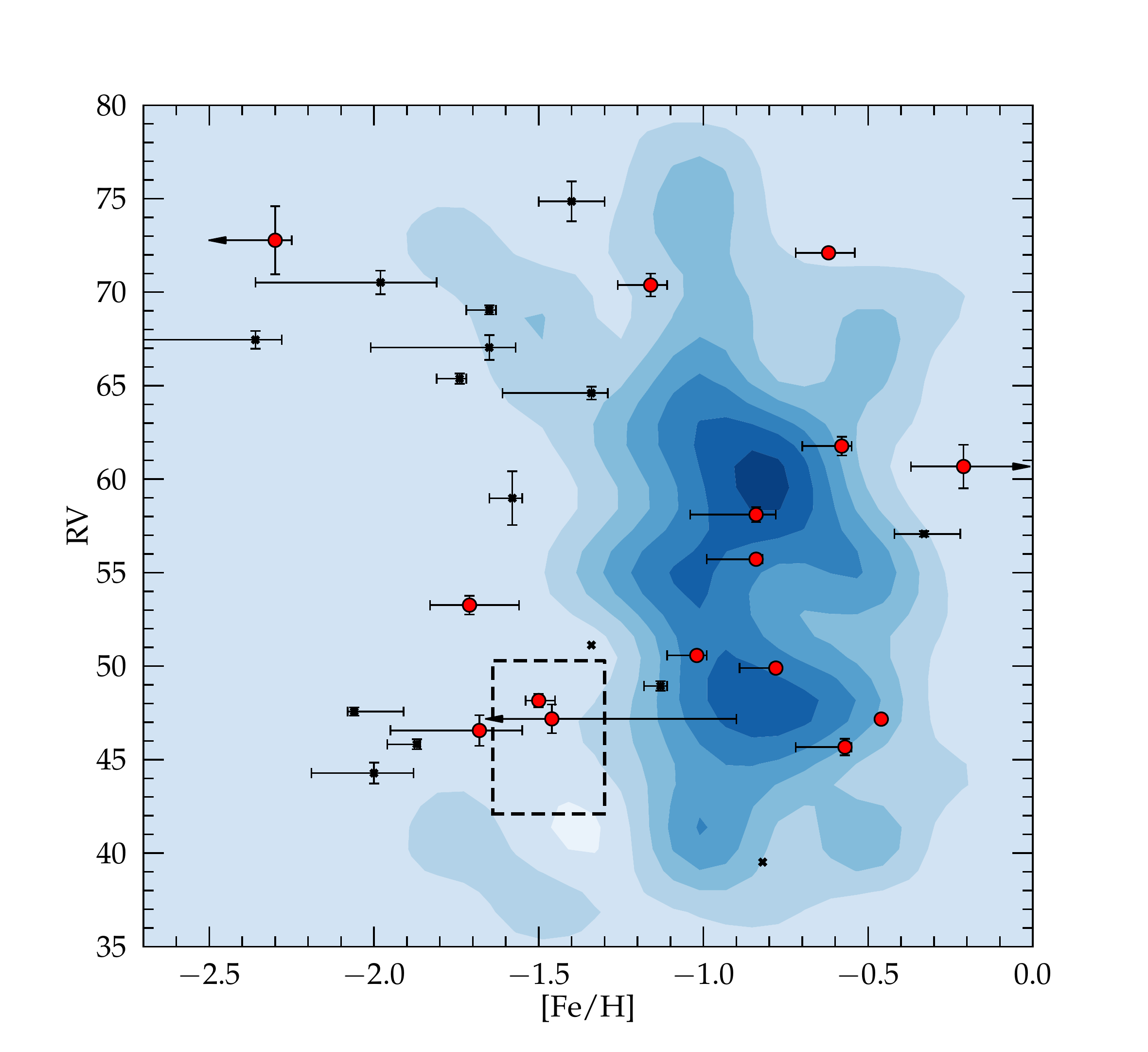}
\caption{Radial velocity and [Fe/H] of our targets stars are overplotted to the field star distribution, visualized as a linearly-scaled density distribution of arbitrary units (blue contours). Stars within the tidal radius of H4 are shown in red. Importantly, three of them group within the parameter space where we expect cluster members (black box). Black symbols are targets outside the tidal radius, which are purposefully selected to be metal-poor and therefore are not representative of the actual field star distribution.}
\label{fig_4_2}
\end{center}
\end{figure}

In Figure~\ref{fig_4_2}, we show the derived [Fe/H] abundances and velocities on top of the underlying field star distribution.
Most of the stars we selected from within the tidal radius of H4 resemble typical properties of field stars in Fornax,
with high metallicities ([Fe/H]$\geq-1.2$) and velocities of $\sim 45 - 70$\,$\mathrm{km\,s^{\rm -1}}$.
From the few outliers, one is very metal-poor ([Fe/H]$\leq -2.0$\,dex) and with high radial velocity. This combination may be typical given that previous studies already detected a trend for more extreme velocities amongst metal-deficient stars in the galaxy (\citealt{Battaglia_06}, \citealt{Hendricks_14b}).
The remaining four all have [Fe/H]$\approx-1.5$, and three of them show almost identical radial velocities of $47.2$, $48.2$, and $46.6$\,$\mathrm{km\,s^{\rm -1}}$. For the last candidate we measure a similar metallicity but a radial velocity about two times larger than the intrinsic dispersion within H4.
Strikingly, two of the stars that fall within the acceptable parameter space are also located closest to the center of H4, which strengthens the assumption that they are members of the cluster. For them, we obtain a membership likelihood of $99.2\pm0.4$\% (target r\_0010) and $97.7\pm1.2$\% (target r\_0016), respectively, but the third star (target b\_0018) with larger $r_{GC}$ is still a $71.0\pm11.0$\% member of H4.
Given a membership likelihood of at least 2-3 sigma for our best candidates, we will assume that these stars are true cluster members in the following chemical analysis and discussion.

The remaining targets outside the tidal radius of H4 have been picked deliberately from the sparse metal-poor tail of the galaxy, and therefore their [Fe/H]-$v$ distribution does not reflect the dominant distribution of the field star population. 

Our chemical analysis yields iron abundances for 30 targets, 15 of which are located within the tidal radius of H4. Of these stars, three ``clump'' around the fiducial cluster properties. Given this small number, our sample does not allow for an individual estimate of the cluster's chemical or dynamical properties, and---in contrast--- we \emph{adopted} these mean cluster parameters from the literature to identify member stars.
From our three likely members, we find a weighted (by error and membership likelihood) mean metallically for our member stars of [Fe/H]$=-1.56$\,dex, slightly lower than the values obtained from integrated-light studies. This may not be surprising when we consider that most of the contaminants in integrated-light spectra are undoubtedly more metal-rich than the cluster itself.
With a similar explanation, our bona-fide members have weighted mean radial velocity of $47.6$\,$\mathrm{km\,s^{\rm -1}}$, marginally higher than the integrated-light estimates. 

\subsection{Alpha-elements}

The alpha-elements in stellar atmospheres analyzed with respect to iron reveal the fraction of SN~Ia contributions to the star forming material and hence are an indicator for the enrichment efficiency of the galactic environment. The [$\mathrm{\alpha}$/Fe] parameter is known to be distinctively different in the field star population of dwarf galaxies compared to the MW.

\begin{figure}[h]
\begin{center}
\includegraphics[width=0.5\textwidth]{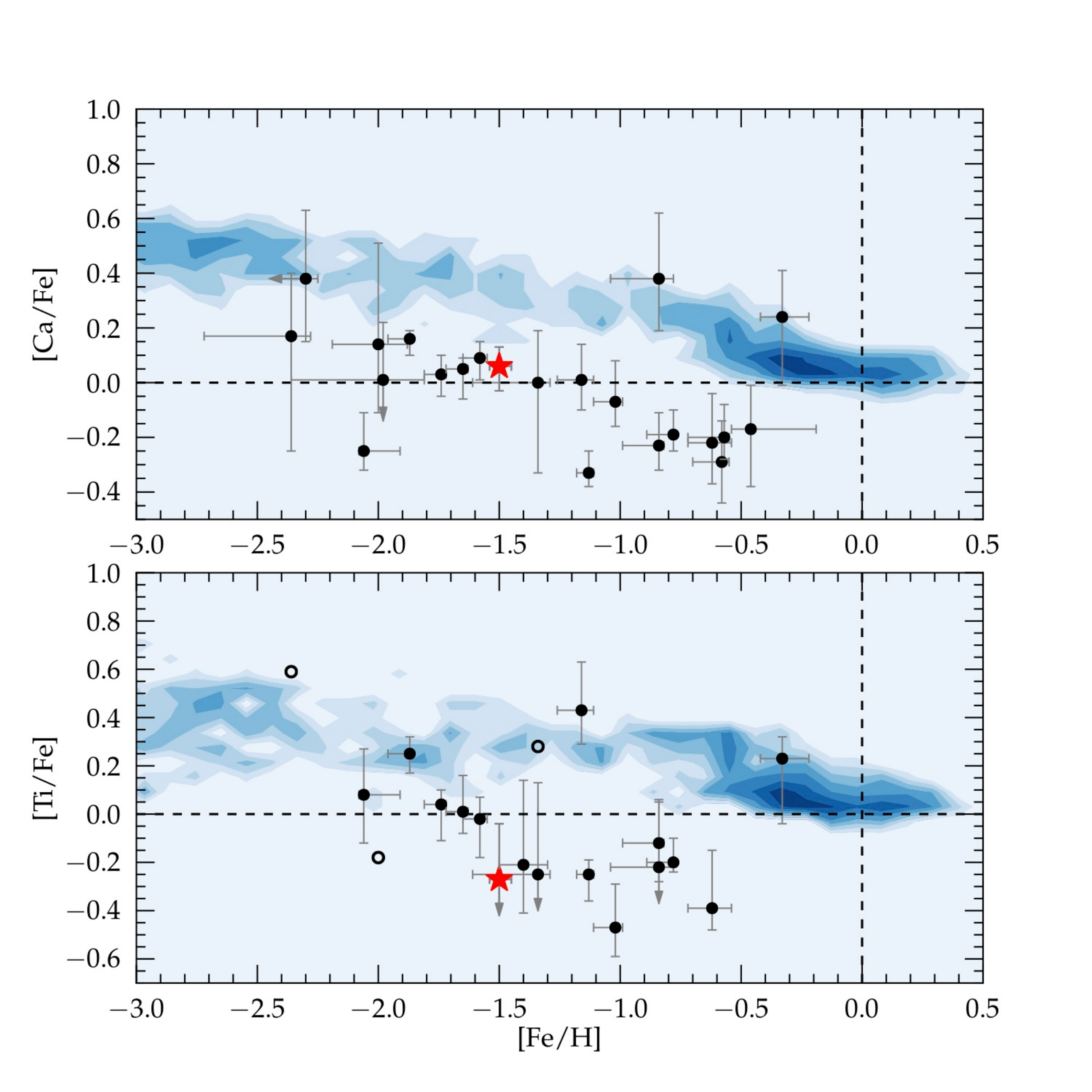}
\caption{Chemical evolution of the alpha-elements Ca (upper panel) and Ti (lower panel) as a function of [Fe/H]. Black dots show the field stars in our sample and the red star highlights the GC H4 star. For visual comparison, the pattern of MW disc and halo stars are shown as a logarithmically-scaled number density distribution of arbitrary units (data from \citealt{Venn_04} and \citealt{Roederer_14}). For both elements, we observe a clear sequence with [Fe/H], and in both cases the H4 member falls on top of this sequence and does not agree with the chemical abundance pattern seen in the MW. Open circles indicate objects for which \textsc{SP\_Ace} could not find upper and lower abundance limits, either due to the low quality of the spectrum or because its chemical parameters fall close to the boundary of the allowed abundance grid.}
\label{fig_4_3}
\end{center}
\end{figure}

In Figure~\ref{fig_4_3} we show the results for the alpha-elements Ca and Ti as a function of the stars' iron abundances.
For both elements, we find a clear sequence with metallicity. While field stars with [Fe/H]$\leq-1.8$ tend to have [$\mathrm{\alpha}$/Fe] above solar level, this ratio drops continuously to clearly sub-solar values for [Fe/H]$\approx -1.3$ and higher. 
Clearly, our stars show an early depletion in both elements compared to the MW halo. This trend does not allow for a knee in the alpha-iron evolution at [Fe/H] significantly higher than $-2.0$\,dex, which confirms previous findings by \citet{Hendricks_14a} and later \citet{Lemasle_14}, and indicates a low chemical enrichment efficiency within the Fornax dSph galaxy.
One star falls out of this general scheme by showing high [$\mathrm{\alpha}$/Fe] for Ca and Ti, although it is the star with the highest [Fe/H] in our sample. By that, it better resembles the characteristics of MW disc stars.
Therefore, it is possible that this star is a Galactic interloper that does not belong to the Fornax dwarf spheroidal.

While two of the H4 members in our sample are too faint to determine reliable alpha-abundances, one star (r\_0010 with 99.2\% membership likelihood) is one of our brightest targets and we can derive precise abundances for Ca, Ti, and, with larger uncertainties, Si.
For this star, we measure low [$\mathrm{\alpha}$/Fe] ratios for all three elements. In detail, we find [Ca/Fe]$=+0.05\substack{+0.09 \\ -0.07}$, [Ti/Fe]$=-0.27\pm{0.23}$, and [Si/Fe]$=-0.35\pm{0.34}$, resulting in an average [$\alpha$/Fe]$=-0.19\pm0.14$. 
Given the high membership likelihood to H4 and the small uncertainties on our abundances, this is a strong indication that H4 is an alpha-depleted GC with a combined [$\mathrm{\alpha}$/Fe] possibly at or around sub-solar level.

To this point, the only existing measurement of alpha-elements in H4 comes from \citet{Larsen_12} and is based on integrated-light spectroscopy of the cluster. These authors find [Ca/Fe]=$+0.13\pm0.07$ and [Ti/Fe]=$+0.12\pm0.05$
, which is somewhat larger than our results from an individual member star.
The discrepancy might be partially explained with the measurement errors of the respective analyses.
Furthermore, it is important to recall the high contamination of more than $20\%$ within the $5\arcsec$ slit, which has been used to obtain the integrated-light spectrum. These authors carefully try to minimize the impact of this contamination by evaluating the spectrum at different positions, and therefore at different cluster radii. Unfortunately, the options to detect and correct for contamination effects are limited, and the contamination fraction does not change by more than 10\% within the central $5\arcsec$ of the cluster. As a consequence, it is difficult to reconstruct the actual underlying population mix because the contaminating field stars show a broad range of ages and metallicities. Therefore, even with a careful approach, a small bias between results obtained from integrated-light and individual stars can be expected. Overall, however, both studies agree that H4 has lower[$\mathrm{\alpha}$/Fe] than similar metallicity Milky Way GC and field stars.

\subsubsection{The full picture: Co-evolution of field stars and GCs in Fornax}

\begin{figure}[t]
\begin{center}
\includegraphics[width=0.5\textwidth]{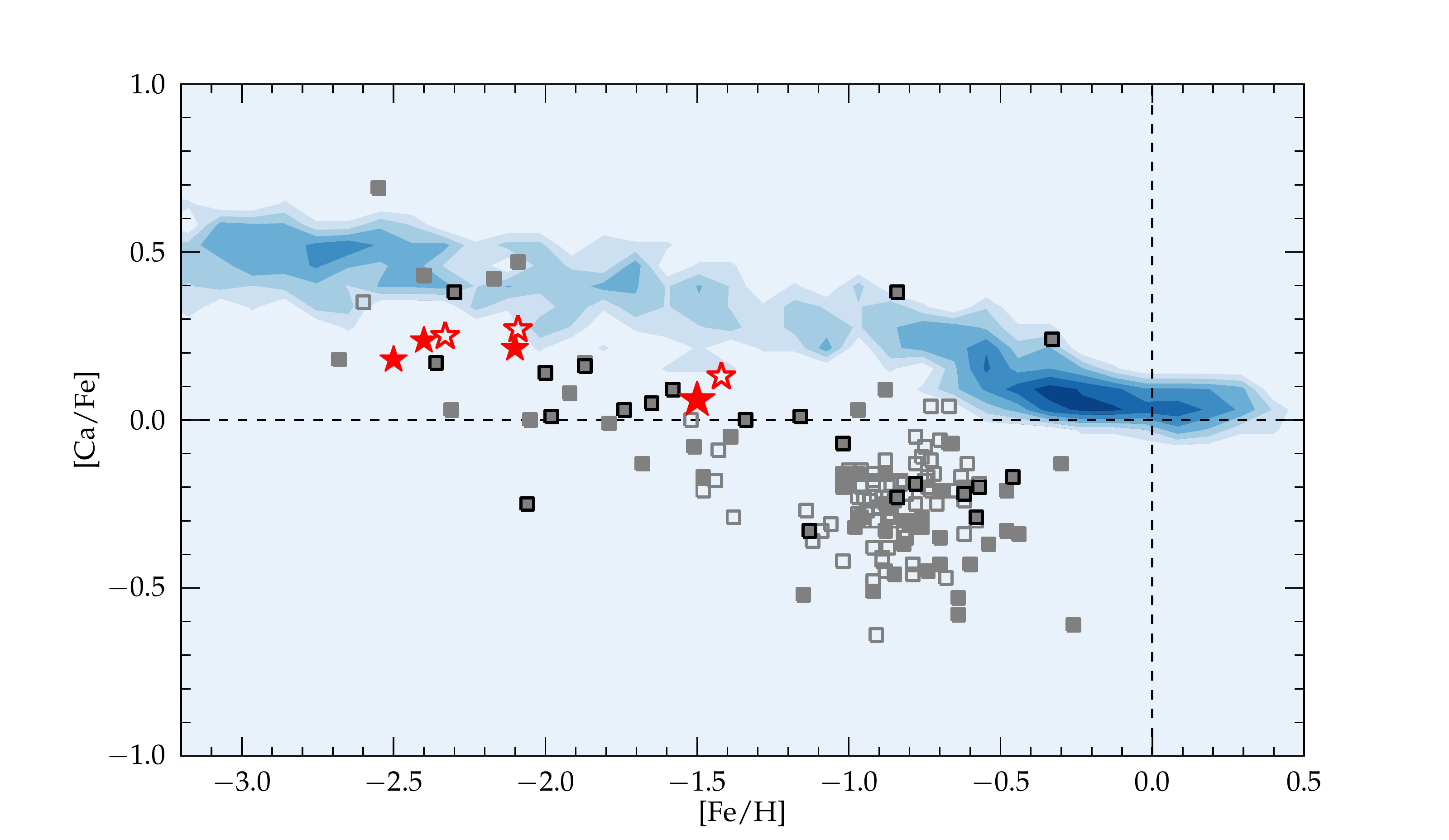}
\caption{Fornax displays a different chemical enrichment pattern from the MW. This is seen in the alpha-element evolution of the field stars \emph{and} the GCs, which show a coupled chemical enrichment. Blue contours show the pattern of MW field stars from \citealt{Venn_04} and \citealt{Roederer_14}. Fornax field stars come from \citet{Letarte_10} (open gray), \citet{Lemasle_14} (filled gray), and this work (filled gray with black edge). Star symbols show the location of the metal-deficient GCs H1, H2, H3 (filled, small red stars; \citealt{Letarte_06}), H3, H4, H5 (open red stars; \citealt{Larsen_12}) and the measurement for H4 presented in this work (filled large red star).}
\label{fig_4_4}
\end{center}
\end{figure}

Important insights can be obtained when the alpha-abundances of all GCs in Fornax are viewed in the context of the field star enrichment in the galaxy.
If we combine literature samples of [Ca/Fe] measurements for field stars (43 stars from \citealt{Lemasle_14}, 85 stars from \citealt{Letarte_10}) with our own sample from this work (21 stars), we obtain a well defined and coherent alpha-element evolution sequence (see Figure~\ref{fig_4_4}). The data clearly show that for [Fe/H] below approximately $-2.0$~dex, field stars are alpha-enhanced and share the typical properties of MW halo stars (see also \citealt{Tafelmeyer_10} for one star at [Fe/H]$=-3.66$ and [Ca/Fe]$=+0.48$). Towards higher metallicities, the [$\mathrm{\alpha}$/Fe] ratio drops and evolves from super-solar to sub-solar values around [Fe/H]$\approx-1.5$.
Although there are no stars in common between the individual field star samples, there is an excellent agreement between the observed properties for all respective metallicities. 

The four metal-poor GCs in Fornax are all moderately enhanced in [$\mathrm{\alpha}$/Fe], and have values comparable to clusters found in the halo of the MW. However, the metal-poor Fornax GCs may lie somewhat below the average plateau found in the MW.
In detail, for three of the metal-deficient clusters in Fornax, [Ca/Fe] has been measured from three individual stars in each cluster by \citet{Letarte_06}. These authors find alpha-enhanced values for all three systems with average values of $+0.18\pm0.09$, $+0.22\pm0.06$, $+0.24\pm0.03$ for H1, H2, and H3, respectively. 
Additionally, \citet{Larsen_12} also derive [Ca/Fe] from integrated-light analysis for the metal-poor GCs H3 and H5 and find $+0.25\pm0.07$ and $+0.27\pm0.05$.

Taking the information from field stars and GCs in Fornax together, two important consequences arise.
First, it is very likely that the cluster H4 does have a significantly lower [$\mathrm{\alpha}$/Fe] abundance ratio than the rest of the GC population in Fornax.
Second, the clusters in Fornax \emph{follow} the [$\mathrm{\alpha}$/Fe] sequence of the field stars and clearly disagree with with the enrichment pattern of the MW halo. 
Therefore, there is strong evidence that the chemical enrichment of GCs and field stars in the Fornax dSph is coupled, and that the clusters trace the chemical signatures of the field star population in their host. We will discuss the consequences of such a scenario for other alpha-depleted GCs in Section~\ref{chapter_discussion_1}.

The evident chemical enrichment pattern also highlights the star formation differences between Fornax and the MW. In the MW, [$\mathrm{\alpha}$/Fe] declines for the field stars but not the GCs, which means that the GCs with, e.g., [Fe/H]$\approx-0.3$ possibly formed before the field stars of the same metallicity. Such a scenario does not appear to be the case with Fornax where the field stars and GCs follow each at metallicities exceeding the location of the [$\mathrm{\alpha}$/Fe] knee.

\subsection{Iron-peak elements}

\begin{figure}[h]
\begin{center}
\includegraphics[width=0.5\textwidth]{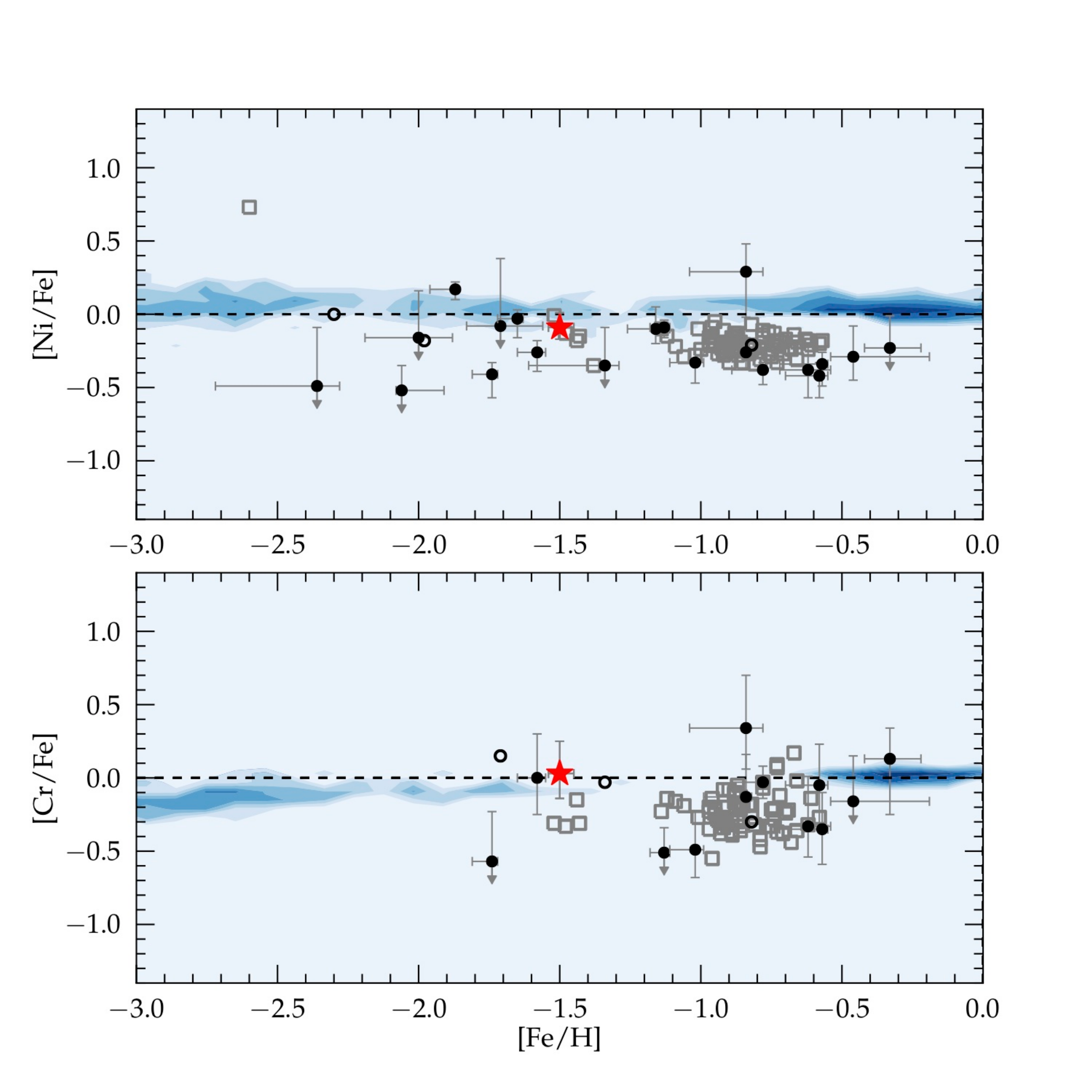}
\caption{[Ni/Fe] and [Cr/Fe] as a function of [Fe/H]. As in Figure~\ref{fig_4_3}, black dots show the field stars in our sample and the red star highlights the GC H4 star. Contours show the pattern of MW disc and halo stars (data from \citealt{Venn_04}, \citealt{Roederer_14}, \citealt{Reddy_03}, and \citealt{Bensby_03}). Open circles indicate objects for which \textsc{SP\_Ace} could not find upper and lower abundance limits. Gray open squares are Fornax field star measurements from \citealt{Letarte_10}.}
\label{fig_4_5}
\end{center}
\end{figure}

It is thought that most iron-peak elements (Sc to Zn: $21 \leq Z \leq 30$, excluding Ti which behaves like an alpha-element) descend from similar nucleosynthetic pathways. While at early times, massive stars and their subsequent SNe~II explosions are the only production resource, SNe~Ia become the dominant contributor later on, when low-mass stars had time to sufficiently evolve. Although they form a common family of elements, the exact formation channels for individual species are not clear, and in most cases there is no simple scaling with the Fe abundance.
For a large fraction of stars in our sample, we obtain abundances for several iron-peak elements to compare with the Galactic trends. The results are shown in Figure~\ref{fig_4_5}

\emph{Nickel.}
The [Ni/Fe] ratio is generally underabundant by $\sim0.2$\,dex compared to solar and MW abundance ratios. Such low abundances have been previously observed in Fornax (\citealt{Letarte_10}, \citealt{Lemasle_14}) and also in Sagittarius (e.g., \citealt{Carretta_10}, \citealt{Sbordone_07}). The LMC displays a larger scatter in Ni above and below solar (\citealt{Pompeia_08}). Our results seem to indicate that the sub-solar [Ni/Fe] abundance ratio, which has been perviously found for stars in satellite systems between [Fe/H]$\approx-0.5$\,dex and [Fe/H]$\approx-1.5$\,dex, continues to even lower metallicities. 

\emph{Chromium.}
Our Cr abundances lie slightly below the solar abundance ratio and the Galactic trend, although few reliable comparison data are available for [Fe/H]$\geq-1.5$\,dex. They agree well, however, with previous measurements of Fornax field stars from \citet{Letarte_10} and \citet{Lemasle_14}, although only a few lines could be measured and the abundances are statistically not very well constrained.

We additionally measure abundances for the iron-peak elements Sc, V, and Co, for which we generally find slightly sub-solar abundance ratios. Unfortunately, there is only very few comparison data available for MW stars, and no previous measurements exist for Fornax. These abundances are therefore not shown in Figure~\ref{fig_4_5}, and they are only listed in Table~\ref{table_a_3} in the Appendix.

\subsection{The age of H4}

The relative age of H4 has been a subject of controversy during the last decades, not least because its detailed chemical composition---and specifically the [$\mathrm{\alpha}$/Fe] abundance of its stars---has not been known. 
Using a relatively clean sample of HST photometry, \citet{Buonanno_99} estimated the relative age of H4 from its photometric offset at different regions in the CMD compared to other clusters in Fornax. From this, the authors found H4 to be $\sim3$~Gyr younger than the other four GCs, which are all coeval and resemble typical old ($\sim12$~Gyr), metal-poor MW halo GCs like M92 (\citealt{Buonanno_98}). However, these results were based on the assumption that \emph{all} Fornax clusters display a similar chemical mixture, in disagreement with our findings.
Later, \citet{Strader_03} used age sensitive---but alpha-insensitive---spectroscopic indices like $H_{\beta}$ and $H_{\gamma}$ to constrain relative ages amongst Fornax clusters and found a similar age for H4 compared to the old systems H2 and H3 (they found, however, indications that H5 is slightly younger).
If H4 is indeed younger compared to the rest of the GC population, it is not clear how Fornax was able to form this cluster several Gyr later than all of the other more metal-poor clusters in the galaxy. This is specifically intriguing when viewed in the context that similarly young Galactic GCs, such as Ruprecht~106 and Palomar~12, are thought to be accreted from satellite dwarf galaxies (\citealt{Brown_97}).

Precise relative age estimates of GCs can be obtained when isochrones are fitted to to the resolved main-sequence turn-off (MSTO) region of the cluster. If the chemical composition and the line-of-sight reddening to the cluster are known, the addition of zero-age horizontal branch (ZAHB) model fitting can enable one to achieve a precision well below 1\,Gyr (see \citealt{VandenBerg_13}).
However, both the position and shape of the MSTO and the HB luminosity are sensitive to the underlying [$\mathrm{\alpha}$/Fe] ratio. This degeneracy causes a systematic bias in the derived ages of several Gyr for cases where the detailed composition of the cluster is not known. 

Provided with a tight constraint on the [$\mathrm{\alpha}$/Fe] abundance in H4, we can now make a new approach to constrain the age of the system and specifically aim to answer the question whether H4 is younger than the remainder of the population in Fornax. To do so, we follow the general procedure described in \citet{VandenBerg_13}. Specifically, we derive synthetic ZAHB loci from the lower envelope of synthetic HB models from \citet{Dotter_07} provided on the Dartmouth-Isochrone database\footnote{{http://stellar.dartmouth.edu/~models/index.html}} and fit them to the observed red horizontal branch of the cluster to set the absolute magnitude scale.
Then, the age of the cluster is determined by the best fitting Dartmouth isochrones (\citealt{Dotter_08}) to the turn-off and subsequent subgiant branch region, after a reddening correction has been applied. To minimize the number of field stars in the CMD, we consider only an area within $r_{GC}\leq15\arcsec$ around the cluster center. 
For all models, we adopt [Fe/H]$=-1.5$ and [$\alpha$/Fe]$=0.0$, and a correction for reddening and extinction of $E(V-I)=0.11$ and $A_{V}=0.248$ (see Section~\ref{chap_chemical_analysis} for details about the reddening).

\begin{figure}[h]
\begin{center}
\includegraphics[width=0.5\textwidth]{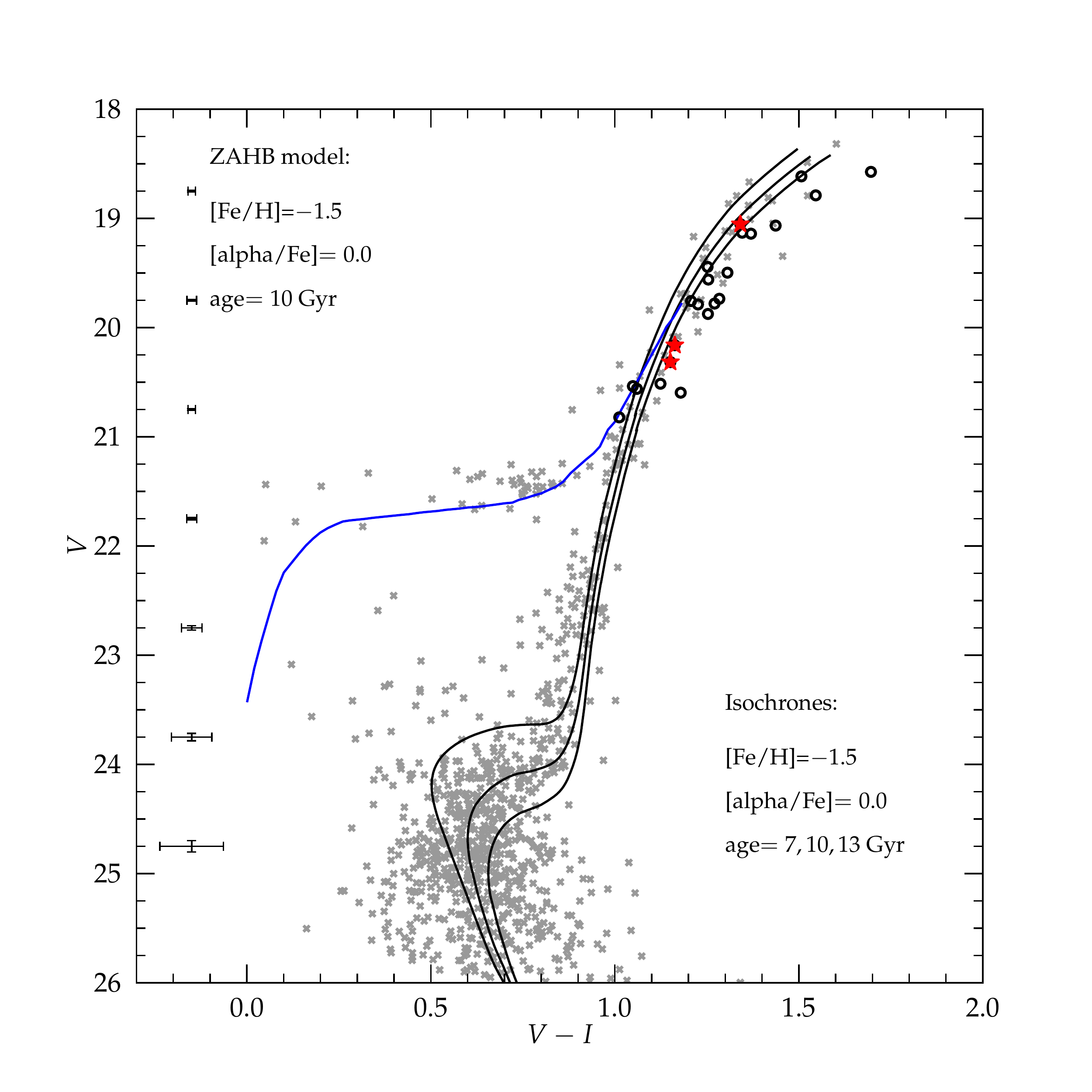}
\caption{Color-magnitude diagram of H4 from our HST photometry. The HST system bands (F555W and F814W) have been transformed to Johnson- Cousins V and I magnitudes (see Section~\ref{chapter_target_selection_1} for details). Only stars from the innermost $15\arcsec$ around the cluster center are shown to minimize the impact of contaminating field stars. Dartmouth Isochrones (\citealt{Dotter_08}) of 7, 10, and 13 Gyr of age and a 10~Gyr ZAHB model are overplotted to the data. The ZAHB luminosity, however, is not sensitive to age in this regime and therefore the model is representative for all three ages compared here. Red stars indicate the position of the three likely H4 members in our sample and black circles are other targets with H4's tidal radius. The isochrones give a best fit for ages around $\sim10$\,Gyr, but uncertainties in reddening and distance to the cluster do not allow a tight age constraint, although the chemical parameters of the cluster are now known.}
\label{fig_4_6}
\end{center}
\end{figure}

In Figure~\ref{fig_4_6} we show the result. From the position of the ZAHB, we find a distance modulus of $\mu_{0}=20.74\pm0.4$, in agreement with previous estimates for the distance to the galaxy itself (e.g., \citealt{Bersier_00}, \citealt{Rizzi_07}, \citealt{Pietrzynski_09}).
However, this number is slightly higher than the results of \citet{Greco_07} who find $\mu_{0}=20.53\pm0.09$ from RR~Lyrae stars in H4, if [Fe/H]$=-1.5$ is adopted.

We obtain the best isochrone fit using an age of $10$~Gyr, which places H4 at a younger age compared to the rest of the population.
However, this result should be interpreted with caution since the uncertainty is at least $\pm1$~Gyr, given the poorly defined MSTO at $V$-band magnitudes around $24.5$ and when we consider the still significant fraction of field stars in the CMD with different chemical signatures and ages than the cluster itself. If we further consider inaccuracies in the reddening \footnote{for H4, the estimates range between $E(B-V)=0.15$ and $0.08$, while Fornax itself has a reddening not higher than $E(B-V)=0.04$ from \citet{Schlegel_98} reddening maps}, an age as old as 12~Gyr, comparable to the other Fornax clusters, seems to be still a conceivable possibility.

\section{Discussion} 
\label{chap_05}

\subsection{Origin of alpha-depleted GCs in the halos of larger galaxies}
\label{chapter_discussion_1}

The evidence that H4 is an alpha-depleted GC makes it one out of only three known clusters to be depleted at this low metallicity. The other known alpha-depleted clusters are Ruprecht~106 in the MW (\citealt{Brown_97}, \citealt{Villanova_13}) and G002 in M31 (\citealt{Colucci_14}).
Importantly, H4 is the only of these clusters which can be clearly associated with a dwarf galaxy, and we may observe in Fornax the first birthplace of a metal-deficient, alpha-depleted GC. The case of Fornax therefore supports speculations which predict similarly depleted clusters in the halo of larger galaxies to originate from dwarf galaxies with low chemical enrichment efficiency (e.g., \citealt{Colucci_14}).

Following this concept, we can use the properties of accreted GCs to learn about their former host systems.
On the one hand, alpha-depleted GCs like Ruprecht~106 and G002 require a host galaxy with a sufficiently low chemical enrichment efficiency to enable significant SNe~Ia contribution for stars already at such low metallicities. Given that a galaxy's mass is likely the main driver of its enrichment efficiency (e.g.,\citealt{Matteucci_90}, \citealt{Tolstoy_09}), the host needs to have stellar masses as low as Fornax or the Sculptor dSph (i.e., a few $10^7\,M_{\sun}$), while a system as massive as the Magellanic Clouds can be ruled out.
On the other hand, recent studies argue that a large fraction of GCs in the halos of the MW and M31 have not formed in-situ, but instead have been accreted from infalling satellite systems (e.g., \citealt{Mackey_04}, \citealt{Leaman_13}, \citealt{Mackey_10}, \citealt{Huxor_11}, \citealt{Elmegreen_12}).
Since in both galaxies the large majority of GCs are uniformly alpha-enhanced over a wide range of iron abundance, some clusters with high [Fe/H] and high [$\mathrm{\alpha}$/Fe]-ratios should be of accreted origin. In contrast to Ruprecht~106 and G002, these clusters require a progenitor system with a higher stellar mass than Fornax in order to prevent a significant contribution of SNe~Ia at these high metallicities.
In a $\Lambda$CDM universe, such high-mass mergers are increasingly unlikely. Consequently, alpha-enhanced accreted GCs necessarily need to originate from only a very small number of disrupted satellite systems.
Strikingly, studying the bifurcated age-metallicity relation amongst disc- and halo GCs in the MW, \citet{Leaman_13} come to a very similar conclusion for both the masses and the number of accreted host satellites carrying GC systems.
Finally, the case of Fornax shows that not all GCs within a dwarf galaxy need to have the same chemical signature. Therefore GCs with different [Fe/H] and [$\mathrm{\alpha}$/Fe] could have originated in the same system.

It seems, there is another lesson to be learned from the observation of alpha-depleted GCs.
There is no evidence, neither from observations nor from theory, about the lower mass limit for a galaxy to be able to form (and hold) own GCs. The search and detection of alpha-depleted clusters can serve as an empirical upper limit. With decreasing [Fe/H] observed in these peculiar clusters, the lower becomes the required star formation efficiency---and hence mass---of the host. Currently, this limit is set by G002, the most metal-deficient alpha-depleted cluster with [Fe/H]$=-1.66$. This cluster consequently requires a host galaxy even smaller than Fornax, and therefore with a maximum mass of $10^7$\,$M_{\sun}$.

\subsection{H4: The nucleus of the Fornax dSph?} 

H4 is located remarkably close to the center of the Fornax dSph. This, together with its distinguishing higher metallicity compared to the other clusters, has fired speculations about whether H4 is in fact the nucleated core of the galaxy, similar to M54 in the Sagittarius dSph galaxy (\citealt{Hardy_02}, \citealt{Strader_03}) or comparable to the suspect accreted nucleus $\omega$~Cen (\citealt{Bekki_03}). If this is the case, it is not self-evident if the properties we observe in H4 can be transferred to classical GC systems in other galaxies.

First, M54 and specifically $\omega$~Cen display a spread in iron. This characteristic cannot be constrained with our data nor from integrated light. We do not observe sufficient individual stars with sufficient chemical precision in order to place a limit on the intrinsic metallicity spread of the system. Integrated light spectroscopy, on the other hand, only provides a cumulative iron abundance while the necessary information about line-strength variation is lost in the doppler-broadened line profile (\citealt{McWilliam_08}). 

Second, M54 is embedded in the very center of its host galaxy. There is also no clear answer to this criterion. If H4 in fact falls on top of the cusp of the field stellar distribution is a matter of debate because of the asymmetry in Fornax' density profile (\citealt{Stetson_98}). This results in a ``chaotic'' behaviour of centroids and inclination angles for elliptical profiles fitted at different radii (\citealt{Demers_94}). While these authors claim to find H4 at the position where the surface density of stars peaks, \citet{Hodge_61} and later \citet{Stetson_98} find an offset between the peak density and the position of H4. If H4 is the nucleated core of Fornax, it should also be measured at the exact same distance. From ZAHB models we find a best fitting distance modulus of $\mu_0=20.74$, which agrees with previous distance measurements for the field star population in Fornax, ranging between $\mu_0=20.65$ (\citealt{Bersier_00}) and $20.87$ (\citealt{Pietrzynski_03}). Given that the actual physical size of the galaxy is between 2 and 3~kpc, and by that several times smaller than the uncertainty on its distance estimations, the existing measurements allow for a placement of H4 right in the center of the galaxy as well as several times outside its tidal perimeter.

Third, M54 moves with the main body of Sagittarius. As outlined in Section~\ref{chapter_membership_selection}, the radial velocity of H4 is determined as precise as $1$\,$\mathrm{km\,s^{\rm -1}}$ from integrated-light spectroscopy, and is found to be distinctively different to the mean galactic motion by $\sim9$\,$\mathrm{km\,s^{\rm -1}}$. This, finally, is evidence that H4 is a classical GC which just coincidentally falls close to the line-of-sight towards the center of Fornax. Better data will be necessary to further constrain the first and the second aspects, and to eventually obtain a final conclusion on the nature of H4.

\subsection{On the discrepancy between photometric and spectroscopic metallicity estimates for H4}

The literature metallicity estimates of H4 derived through photometric and spectroscopic methods have yielded inconsistent results: \citet{Buonanno_99} estimates a metallicity $\leq-2.0$\,dex based on the slope of the RGB as compared to the other Fornax clusters, a result which has been  confirmed later by \citet{dAntona_13}. In contrast, spectroscopic measurements consistently yield values around $\sim -1.4$\,dex (e.g., \citealt{Strader_03}, \citealt{Larsen_12})--a discrepancy too large to be explained solely by measurement errors. However, while spectroscopic iron-line analyses yield [Fe/H], broad-band photometric approaches are sensitive to the total mass fraction of all metals ($Z$), which determines the stellar structure and hence the cluster fiducial sequence in a CMD. Therefore, it is possible that the low alpha-abundance ratio we find for H4 is responsible for the discrepancy between photometric and spectroscopic metallicity estimates.

In fact, we find a ratio of metallicity mass fraction between the GCs H4 and, e.g., H2 $Z_{\mathrm{H4}}/Z_{\mathrm{H2}}\approx3.2$, when we assume [Fe/H]$=-1.5$\,dex for H4 and $-2.1$\,dex for H2, and enhanced alpha-abundances (including oxygen) to [$\mathrm{\alpha}$/Fe]$=+0.4$\,dex for both clusters. The metallicity mass fraction ratio decreases to only $\approx1.5$, and hence to about half of the original discrepancy, when we instead adopt solar-like alpha-abundance ratios for H4.
Alternatively, for a cluster with H4's iron abundance, the absolute alteration in $Z$ caused by a change from [$\mathrm{\alpha}$/Fe]$=+0.4$\,dex to solar, amounts to $\Delta Z = 5\times10^{-4}$ when we apply the standard solar abundance scale from \citet{Asplund_09}. This number is about equivalent to a shift in [Fe/H] from $-1.5$\,dex to $-1.8$\,dex when [$\mathrm{\alpha}$/Fe] is kept constant.

Thus, at least part of the discrepancy between photometric and spectroscopic abundance determinations for H4 likely stems from the low [$\mathrm{\alpha}$/Fe] ratio in H4 compared to the other clusters, and from the fact that spectroscopic iron abundances have been compared to features that are more sensitive to the combined mass fraction of elements heavier than helium.

\subsection{Insights from field star evolution at different galactocentric radii}

Fornax is one of the best studied of all dwarf galaxies, and the detailed chemical properties of field stars have been an issue in a series of recent papers (\citealt{Letarte_10}, \citealt{Hendricks_14a}, \citealt{Lemasle_14}). It is also the only galaxy where the combined datasets cover a large fraction of its radial extent. Specifically, \citet{Hendricks_14a} and \citet{Lemasle_14} provide the alpha-evolution of stars at $r\approx0.6\degree$ (compared to a tidal radius of $\sim1\degree$), while the current work, for the first time, yield similar information for the very central area at $r\leq0.2\degree$. 

\begin{figure}[h]
\begin{center}
\includegraphics[width=0.5\textwidth]{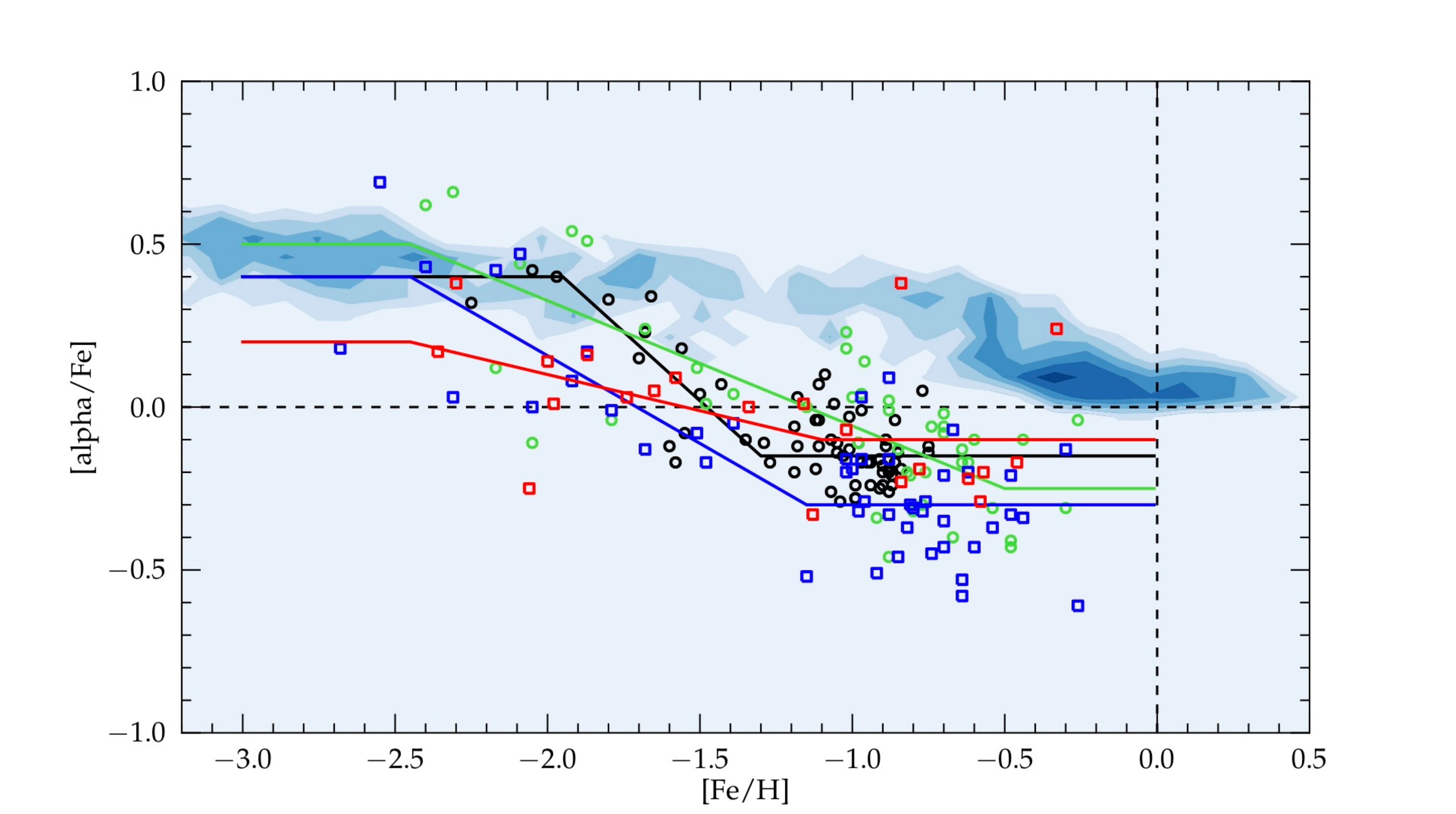}
\caption{Fornax displays very similar chemical enrichment signatures at different radial positions within the galaxy. The evolution of individual alpha-elements is identical within the limited precision of the data. In detail, black circles are [Mg/Fe] measured at $r\approx0.6\degree$ from \citet{Hendricks_14a}, green circles and blue squares show [Mg/Fe] and [Ca/Fe] measured at $r\approx0.6\degree$ from \citet{Lemasle_14}, and red squares are the [Ca/Fe] values at $r\approx0.2\degree$ from this work. The lines are toy model fits as described in the text to the samples with corresponding color. The contours show the evolution of MW field stars for comparison, averaged for Mg and Ca.}
\label{fig_5_1}
\end{center}
\end{figure}

In Figure~\ref{fig_5_1}, we fit a simple step function to the alpha-evolution sequence of all literature samples, similarly to what has been done in \citet{Cohen_10}, or \citet{Hendricks_14a}. This toy model has no physical motivation, and is only designed to estimate the position of the knee and the two plateau values of [$\mathrm{\alpha}$/Fe] for both high and low [Fe/H]. When we compare the alpha-evolution at different radial positions in this naive way, we find them to follow essentially the same sequence, which also agrees with the visual impression of the data.

This mutual agreement between the different samples confirms a very slow chemical enrichment in Fornax, seen as a metal-poor knee.
Moreover, this means that the chemical enrichment efficiency in the center and the outskirts of Fornax had to be similar, at least at early times.
This fact is a little surprising, if one considers that the enrichment efficiency comprises the star formation efficiency on the one side, and the ability to retain the processed stellar yields on the other. Both factors are theoretically sensitive to the density of the ISM and the depth of the gravitational potential, which in turn are both a function of radius.

It is well established that stars in the center of Fornax are of a significantly higher average metallicity than in the outer parts (e.g., \citealt{Battaglia_06}). The consequence of the above considerations could be that the inner area did not undergo a \emph{faster} chemical evolution, but rather experienced a \emph{longer} star formation history, which eventually caused the observed radial metallicity gradient within the galaxy.

\section{Summary} 
\label{chap_06}

In this paper, we presented radial velocities and high-resolution chemical abundances for Fe, several alpha-elements (Ca, Ti, and Si), and iron-peak elements (V, Cr, Co, Ni) for an individual member star of the peculiar GC H4 and 27 additional stars in the surrounding field at the center of the Fornax dwarf spheroidal galaxy. H4 is difficult to observe due to its central position within the galaxy with severe contamination higher than 30\% in the resolved area of the cluster. 
Our sample has been selected from HST photometry, where we carefully avoided blended stars in a seeing limited scenario.
We obtain the first detailed chemical analysis for an individual member star in H4, which we put into context to the chemical signatures of field stars in Fornax from this work and from previous studies.
The main results are summarized below:

\begin{itemize}
\item{Our field star sample cover a wide range in metallicity with $-2.3\leq \mathrm{[Fe/H]} \leq -0.4$, and 
show a distinct change in [$\alpha$/Fe] as a function of [Fe/H]. In detail, stars are alpha-enhanced at low metallicity where they show similar properties to metal-deficient MW field stars. With increasing [Fe/H], they follow a clear sequence towards sub-solar [$\mathrm{\alpha}$/Fe] ratios.}

\item{The observed field star sequence is in good agreement with previous observations made in this galaxy and does not allow for a knee in the alpha-evolution significantly higher than [Fe/H]$\sim-2.0$\,dex, indicative of a low star formation efficiency in Fornax.}

\item{By comparing the alpha-evolution of different field star samples from literature, it is possible to put a tentative constraint on radial chemical enrichment variations within Fornax. We do not find any significant variation for the evolution of a given species with [Fe/H] and therefore speculate that such variations, if existent, have to be small.}

\item{We obtain precise chemical abundances for one star with a $99.2$\% membership likelihood to the GC H4 and find low alpha-abundances of [Ca/Fe]$=+0.05\pm0.08$, [Ti/Fe]$=-0.27\pm0.23$, and [Si/Fe]$=-0.35\pm0.34$, resulting in an average [$\alpha$/Fe]$=-0.19$. 
This makes H4 one out of only three metal-deficient GCs known to be alpha-depleted. 
Moreover, Fornax becomes the first observed birthplace of such peculiar clusters, which can also be found in the halos of larger galaxies, supporting speculations that these clusters are accreted from now disrupted satellite systems.}

\item{Considered together, the GC population in Fornax follows the chemical signature of the field stars and explicitly disagrees with the properties of Milky Way stars and GCs. We therefore conclude that the chemical enrichment of field and clusters in Fornax is coupled and determined by the properties of the common host galaxy.}

\item{If the chemical enrichment signatures of a galaxy as imprinted in the alpha-evolution of its field stars are inherited to its GC population, we can draw inferences from the chemical properties of accreted GCs about their unknown satellite hosts.
Following this concept, the alpha-depleted, metal-deficient clusters Ruprecht~106 in the Milky Way and G002 in M31 require a host galaxy similar to, or smaller than Fornax, and consequently with a stellar mass of $\sim 10^7 M_{\sun}$. In contrast, alpha-enhanced GCs with [Fe/H]$\sim-1$, if accreted, require more massive birth places with masses $\sim10^8M_{\sun}$ or more. This argument, however, relies on the assumption that mass is the main parameter to determine the chemical enrichment properties of a galaxy. If, however, other parameters like environmental interactions also play an important role, it becomes more difficult to reconstruct the host galaxies properties in such a way.}
\end{itemize}

Finally, it is important to emphasize that a large portion of our findings and their interpretation rely on the properties of only one member star of H4 (two more likely candidates in our sample proved too faint for a detailed chemical analysis).
Although the analyzed star is almost certainly a member of the cluster, and despite the robust abundance measurement we could perform on its spectrum, it is possible that its properties are not representative of the average cluster chemical composition.
Clearly, in order to confirm the results related to H4 and the subsequent conclusions, it would be desirable to obtain chemical information from more individual members of this cluster in the future.

\begin{acknowledgements}
We thank Charli Sakari, Hans-G\"unter Ludwig, and Morgan Fouesneau for helpful discussions. The HST data used in this paper were obtained from the Mikulski Archive for Space Telescopes (MAST). STScI is operated by the Association of Universities for Research in Astronomy, Inc., under NASA contract NAS5-26555. This publication makes use of data products from the Two Micron All Sky Survey, which is a joint project of the University of Massachusetts and the Infrared Processing and Analysis Center/California Institute of Technology, funded by NASA and the National ScienceFoundation.
BH, MF, and AK acknowledge the German Research Foundation (DFG) for funding from Emmy-Noether grant Ko 4161/1. This work was in part supported by Sonderforschungsbereich SFB 881 "The Milky Way System" (subproject A5) of the DFG. C.I.J. acknowledges support through the Clay Fellowship administered by the Smithsonian Astrophysical Observatory.
\end{acknowledgements}

\begin{appendix}

\section{Abundances and velocities for Fornax field stars and likely members of H4}
\label{Appendix}

\begin{table*}[htb]
\centering
\caption{Basic parameters -- positions, photometry, and quality of the spectra }
\begin{tabular}{lllcccccccc}
\hline
\hline
Star ID	 & \ \ \ \ \ \ \ \ $\alpha$	 & \ \ \ \ \ \ \ \ \ \ \ \ \  $\delta$	 & Group	& $V$	 & $\sigma V$	 & $V-I$	 & $\sigma (V-I)$	 & $m_{sep}$	 & $S/N$	 & $\Delta\lambda_{eff}$[\AA] \\
\hline
r\_0006	 &  2h40m10.15s	 & -34d31m48.83s	 &A	 & 18.892	 & 0.007	 & 1.260	 & 0.009	 & 10.01	 & 23.8	 & 301.6	 \\
r\_0007	 &  2h40m9.19s	 & -34d32m59.89s	 &A	 & 19.540	 & 0.010	 & 1.116	 & 0.013	 & 6.87	   	 & 15.9	 & 393.9	 \\
r\_0008	 &  2h40m5.51s	 & -34d32m42.86s	 &A	 & 18.326	 & 0.005	 & 1.586	 & 0.006	 & 8.16	     & 40.2	 & 439.2	 \\
r\_0009	 &  2h40m8.84s	 & -34d32m47.83s	 &A	 & 19.249	 & 0.008	 & 1.196	 & 0.011	 & 6.82	     & 18.0	 & 214.2	 \\
r\_0010\tablefootmark{*}	 &  2h40m7.69s	 & -34d32m0.92s	 &A	 & 18.805	 & 0.007	 & 1.231	 & 0.009	 & 5.95	     & 33.7	 & 345.7	 \\
r\_0011	 &  2h40m8.30s	 & -34d32m37.28s	 &A	 & 18.818	 & 0.007	 & 1.327	 & 0.009	 & 8.49	     & 27.4	 & 450.6	 \\
r\_0014	 &  2h39m41.20s	 & -34d34m50.09s	 &B	 & 18.242	 & $-$	 & 1.260	 & $-$	 & $-$	 & 33.4	 & 434.7	 \\
r\_0016	 &  2h40m6.63s	 & -34d32m25.80s	 &A	 & 20.068	 & 0.004	 & 1.041	 & 0.005	 & 4.70	     & 13.8	 & 175.3	 \\
r\_0017	 &  2h40m7.22s	 & -34d31m44.94s	 &A	 & 19.488	 & 0.003	 & 1.174	 & 0.004	 & 4.75	     & 17.3	 & 304.7	 \\
r\_0018	 &  2h40m7.07s	 & -34d31m25.79s	 &A	 & 19.196	 & 0.008	 & 1.142	 & 0.011	 & 4.73	     & 19.4	 & 379.4	 \\
r\_0019	 &  2h39m31.50s	 & -34d46m45.12s	 &B	 & 18.082	 & $-$	 & 1.020	 & $-$	 & $-$	 & 34.4	 & 423.2	 \\
r\_0020	 &  2h39m4.80s	 & -34d41m31.49s	 &B	 & 18.422	 & $-$	 & 1.230	 & $-$	 & $-$	 & 43.8	 & 463.3	 \\
r\_0021	 &  2h39m12.29s	 & -34d44m38.54s	 &B	 & 17.912	 & $-$	 & 1.240	 & $-$	 & $-$	 & 39.5	 & 478.8	 \\
r\_0022	 &  2h40m4.01s	 & -34d32m13.60s	 &A	 & 19.312	 & 0.009	 & 1.144	 & 0.012	 & 8.84	     & 21.3	 & 378.0	 \\
r\_0024	 &  2h40m4.46s	 & -34d31m38.64s	 &A	 & 19.532	 & 0.010	 & 1.161	 & 0.013	 & 5.62	     & 18.5	 & 360.6	 \\
b\_0001	 &  2h39m11.42s	 & -34d29m22.27s	 &B	 & 18.012	 & $-$	 & 1.320	 & $-$	 & $-$	 & 16.6	 & 277.0	 \\
b\_0003	 &  2h39m39.90s	 & -34d43m7.21s	 &B	 & 18.362	 & $-$	 & 1.110	 & $-$	 & $-$	 & 51.1	 & 375.9	 \\
b\_0005	 &  2h39m36.30s	 & -34d51m25.99s	 &B	 & 18.062	 & $-$	 & 1.710	 & $-$	 & $-$	 & 19.5	 & 379.4	 \\
b\_0007	 &  2h40m5.97s	 & -34d32m0.89s	 &A	 & 18.367	 & 0.006	 & 1.397	 & 0.007	 & 7.56	     & 38.1	 & 403.0	 \\
b\_0008	 &  2h40m6.09s	 & -34d31m38.28s	 &A	 & 18.883	 & 0.007	 & 1.236	 & 0.009	 & 4.22	     & 20.3	 & 311.7	 \\
b\_0009	 &  2h39m55.38s	 & -34d45m56.05s	 &B	 & 18.282	 & $-$	 & 1.480	 & $-$	 & $-$	 & 26.9	 & 355.3	 \\
b\_0010	 &  2h39m53.79s	 & -34d38m55.50s	 &B	 & 18.322	 & $-$	 & 1.160	 & $-$	 & $-$	 & 29.4	 & 175.3	 \\
b\_0011	 &  2h39m52.62s	 & -34d45m44.60s	 &B	 & 17.982	 & $-$	 & 1.290	 & $-$	 & $-$	 & 55.5	 & 426.4	 \\
b\_0015	 &  2h40m5.09s	 & -34d35m41.75s	 &B	 & 18.332	 & $-$	 & 1.160	 & $-$	 & $-$	 & 51.6	 & 281.2	 \\
b\_0016	 &  2h40m2.73s	 & -34d38m30.05s	 &B	 & 18.162	 & $-$	 & 1.260	 & $-$	 & $-$	 & 33.8	 & 352.0	 \\
b\_0017	 &  2h39m57.10s	 & -34d49m7.03s	 &B	 & 18.412	 & $-$	 & 1.080	 & $-$	 & $-$	 & 27.8	 & 372.3	 \\
b\_0018	 &  2h40m10.92s	 & -34d32m4.31s	 &A	 & 19.913	 & 0.004	 & 1.053	 & 0.005	 & 6.50	     & 11.1	 & 277.7	 \\
b\_0019	 &  2h40m11.20s	 & -34d31m46.78s	 &A	 & 19.627	 & 0.010	 & 1.143	 & 0.013	 & 7.20	     & 11.1	 & 236.1	 \\
b\_0020	 &  2h40m21.48s	 & -34d25m57.43s	 &B	 & 18.402	 & $-$	 & 1.320	 & $-$	 & $-$	 & 13.2	 & 175.0	 \\
b\_0023	 &  2h40m30.22s	 & -34d38m53.81s	 &B	 & 18.142	 & $-$	 & 1.210	 & $-$	 & $-$	 & 37.1	 & 447.7	 \\\hline

\end{tabular}
\tablefoot{``Group'' indicates whether the star has been selected within the tidal radius of H4 (A), or whether it is a field star from the central part of Fornax (B). For stars of category A, the seperation index $m_{sep}$ is an indicator for the flux contamination of a spectrum, as described in Section~\ref{chapter_target_selection_1} of the main article. The S/N is per pixel of the degraded spectra as they are supplied to \textsc{SP\_Ace}. The effective wavelength range $\Delta\lambda_{eff}$ gives the final wavelength range which is used in the chemical analysis after critical regions flagged in various pixel masks have been removed. \tablefoottext{*}{The star with ID r\_0010 is the target with 99.2\% H4 membership probability.}}
\label{table_1}  

\end{table*}

\begin{table*}[htb]
\centering
\caption{Chemical abundances -- [Fe/H], and alpha-elements}
\begin{tabular}{lccccccccccc}
\hline
\hline

Star ID	 & $v$	 & $\sigma v$	 & [Fe/H]	 & [Fe/H]$_{\mathrm{+}}$	 & [Fe/H]$_{\mathrm{-}}$	 & [Ca/H]	 & [Ca/H]$_{+}$	 & [Ca/H]$_{-}$	 & [Ti/H]	 & [Ti/H]$_{+}$	 & [Ti/H]$_{-}$	  \\\hline
r\_0006	 & 53.26	 & 0.50	 & $-1.71$	 & $-1.83$	 & $-1.56$	 & $-$	 & $-$	 & $-$	 & $-$	 & $-$	 & $-$	  \\
r\_0007	 & 70.38	 & 0.61	 & $-1.16$	 & $-1.26$	 & $-1.11$	 & $-1.15$	 & $-1.26$	 & $-1.02$	 & $-0.73$	 & $-0.87$	 & $-0.53$	  \\
r\_0008	 & 72.10	 & 0.20	 & $-0.62$	 & $-0.72$	 & $-0.54$	 & $-0.84$	 & $-0.99$	 & $-0.66$	 & $-1.01$	 & $-1.10$	 & $-0.77$	  \\
r\_0009	 & 58.10	 & 0.40	 & $-0.84$	 & $-1.04$	 & $-0.78$	 & $-0.46$	 & $-0.65$	 & $-0.22$	 & $-1.06$	 & $-$	 & $-0.78$	  \\
r\_0010\tablefootmark{*}	 & 48.16	 & 0.36	 & $-1.50$	 & $-1.54$	 & $-1.45$	 & $-1.44$	 & $-1.53$	 & $-1.37$	 & $-1.77$	 & $-$	 & $-1.54$	  \\
r\_0011	 & 49.89	 & 0.28	 & $-0.78$	 & $-0.89$	 & $-0.77$	 & $-0.97$	 & $-1.03$	 & $-0.88$	 & $-0.98$	 & $-1.02$	 & $-0.88$	  \\
r\_0014	 & 48.94	 & 0.25	 & $-1.13$	 & $-1.18$	 & $-1.11$	 & $-1.46$	 & $-1.51$	 & $-1.38$	 & $-1.38$	 & $-1.49$	 & $-1.32$	  \\
r\_0016	 & 46.56	 & 0.82	 & $-1.68$	 & $-1.95$	 & $-1.55$	 & $-$	 & $-$	 & $-$	 & $-$	 & $-$	 & $-$	  \\
r\_0017	 & 47.17	 & 2.12	 & $-0.46$	 & $-0.54$	 & $-0.19$	 & $-0.63$	 & $-0.84$	 & $-0.47$	 & $-$	 & $-$	 & $-$	  \\
r\_0018	 & 72.78	 & 1.83	 & $-2.30$	 & $-$	 & $-2.25$	 & $-1.92$	 & $-2.15$	 & $-1.67$	 & $-$	 & $-$	 & $-$	  \\
r\_0019	 & 64.60	 & 0.34	 & $-1.34$	 & $-1.61$	 & $-1.29$	 & $-1.34$	 & $-1.67$	 & $-1.15$	 & $-1.59$	 & $-$	 & $-1.21$	  \\
r\_0020	 & 58.98	 & 1.44	 & $-1.58$	 & $-1.65$	 & $-1.55$	 & $-1.49$	 & $-1.57$	 & $-1.43$	 & $-1.60$	 & $-1.76$	 & $-1.51$	  \\
r\_0021	 & 65.38	 & 0.27	 & $-1.74$	 & $-1.81$	 & $-1.72$	 & $-1.71$	 & $-1.79$	 & $-1.64$	 & $-1.70$	 & $-1.85$	 & $-1.64$	  \\
r\_0022	 & 55.72	 & 0.28	 & $-0.84$	 & $-0.99$	 & $-0.82$	 & $-1.07$	 & $-1.16$	 & $-0.95$	 & $-0.96$	 & $-1.12$	 & $-0.79$	  \\
r\_0024	 & 61.77	 & 0.50	 & $-0.58$	 & $-0.70$	 & $-0.55$	 & $-0.87$	 & $-1.02$	 & $-0.72$	 & $-1.20$	 & $-$	 & $-1.01$	  \\
b\_0001	 & 74.85	 & 1.07	 & $-1.40$	 & $-1.50$	 & $-1.30$	 & $-1.94$	 & $-2.05$	 & $-1.69$	 & $-1.61$	 & $-1.81$	 & $-1.26$	  \\
b\_0003	 & 67.45	 & 0.48	 & $-2.36$	 & $-2.72$	 & $-2.28$	 & $-2.19$	 & $-2.61$	 & $-1.96$	 & $-1.77$	 & $-$	 & $-$	  \\
b\_0005	 & 39.52	 & 0.48	 & $-0.82$	 & $-$	 & $-$	 & $-$	 & $-$	 & $-$	 & $-$	 & $-$	 & $-$	  \\
b\_0007	 & 50.57	 & 0.19	 & $-1.02$	 & $-1.11$	 & $-0.99$	 & $-1.09$	 & $-1.18$	 & $-0.94$	 & $-1.49$	 & $-1.61$	 & $-1.31$	  \\
b\_0008	 & 45.67	 & 0.44	 & $-0.57$	 & $-0.72$	 & $-0.55$	 & $-0.77$	 & $-0.85$	 & $-0.65$	 & $-$	 & $-$	 & $-$	  \\
b\_0009	 & 51.13	 & 0.33	 & $-1.34$	 & $-$	 & $-$	 & $-$	 & $-$	 & $-$	 & $-1.06$	 & $-$	 & $-$	  \\
b\_0010	 & 44.28	 & 0.56	 & $-2.00$	 & $-2.19$	 & $-1.88$	 & $-1.86$	 & $-2.11$	 & $-1.49$	 & $-2.18$	 & $-$	 & $-1.68$	  \\
b\_0011	 & 47.57	 & 0.22	 & $-2.06$	 & $-2.08$	 & $-1.91$	 & $-2.31$	 & $-2.38$	 & $-2.17$	 & $-1.98$	 & $-2.18$	 & $-1.79$	  \\
b\_0015	 & 45.82	 & 0.26	 & $-1.87$	 & $-1.96$	 & $-1.86$	 & $-1.71$	 & $-1.77$	 & $-1.68$	 & $-1.62$	 & $-1.70$	 & $-1.55$	  \\
b\_0016	 & 69.05	 & 0.24	 & $-1.65$	 & $-1.72$	 & $-1.63$	 & $-1.60$	 & $-1.71$	 & $-1.56$	 & $-1.64$	 & $-1.73$	 & $-1.49$	  \\
b\_0017	 & 70.52	 & 0.63	 & $-1.98$	 & $-2.36$	 & $-1.81$	 & $-1.97$	 & $-$	 & $-1.76$	 & $-$	 & $-$	 & $-$	  \\
b\_0018	 & 47.18	 & 0.77	 & $-1.46$	 & $-$	 & $-0.90$	 & $-$	 & $-$	 & $-$	 & $-$	 & $-$	 & $-$	  \\
b\_0019	 & 60.68	 & 1.16	 & $-0.21$	 & $-0.37$	 & $-$	 & $-1.20$	 & $-$	 & $-0.20$	 & $-0.88$	 & $-$	 & $-0.36$	  \\
b\_0020	 & 67.04	 & 0.66	 & $-1.65$	 & $-2.01$	 & $-1.57$	 & $-1.84$	 & $-$	 & $-1.57$	 & $-$	 & $-$	 & $-$	  \\
b\_0023	 & 57.07	 & 0.15	 & $-0.33$	 & $-0.42$	 & $-0.22$	 & $-0.09$	 & $-0.34$	 & $0.08$	 & $-0.10$	 & $-0.37$	 & $-0.01$	  \\\hline

\end{tabular}
\tablefoot{\tablefoottext{*}{The star with ID r\_0010 is the target with 99.2\% H4 membership probability.}}
\label{table_2}  

\end{table*}

\begin{sidewaystable*}[htb]
\centering
\caption{Chemical abundances -- iron-peak elements}
\begin{tabular}{lcccccccccccc}
\hline
\hline

Star ID	 & [V/H]	 & [V/H]$_{+}$	 & [V/H]$_{-}$	 & [Cr/H]	 & [Cr/H]$_{+}$	 & [Cr/H]$_{-}$	 & [Co/H]	 & [Co/H]$_{+}$ & [Co/H]$_{-}$	 & [Ni/H]	 & [Ni/H]$_{+}$	 & [Ni/H]$_{-}$	    \\\hline
r\_0006	 & $-1.54$	 & $-2.00$	 & $-1.33$	 & $-1.56$	 & $-$	 & $-1.04$	 & $-1.09$	 & $-1.33$	 & $-$	 & $-1.79$	 & $-$	 & $-1.33$	   \\
r\_0007	 & $-0.91$	 & $-1.01$	 & $-0.70$	 & $-$	 & $-$	 & $-$	 & $-0.94$	 & $-1.10$	 & $-0.73$	 & $-1.26$	 & $-1.36$	 & $-1.11$	   \\
r\_0008	 & $-0.92$	 & $-1.00$	 & $-0.68$	 & $-0.95$	 & $-1.16$	 & $-0.61$	 & $-0.96$	 & $-1.10$	 & $-0.73$	 & $-1.00$	 & $-1.19$	 & $-0.83$	   \\
r\_0009	 & $-0.25$	 & $-0.58$	 & $-$	 & $-0.50$	 & $-0.78$	 & $-0.14$	 & $-1.06$	 & $-$	 & $-0.82$	 & $-0.55$	 & $-0.86$	 & $-0.36$	   \\%\hline
r\_0010\tablefootmark{*}	 & $-1.65$	 & $-1.73$	 & $-1.49$	 & $-1.47$	 & $-1.64$	 & $-1.25$	 & $-1.83$	 & $-1.95$	 & $-1.70$	 & $-1.59$	 & $-1.67$	 & $-1.49$	   \\%\hline
r\_0011	 & $-1.12$	 & $-1.16$	 & $-0.99$	 & $-0.81$	 & $-0.92$	 & $-0.70$	 & $-1.04$	 & $-1.13$	 & $-0.94$	 & $-1.16$	 & $-1.26$	 & $-1.11$	   \\
r\_0014	 & $-1.41$	 & $-1.49$	 & $-1.32$	 & $-1.64$	 & $-$	 & $-1.47$	 & $-1.21$	 & $-1.29$	 & $-1.13$	 & $-1.22$	 & $-1.29$	 & $-1.17$	   \\
r\_0016	 & $-$	 & $-$	 & $-$	 & $-$	 & $-$	 & $-$	 & $-1.21$	 & $-1.68$	 & $-0.89$	 & $-1.44$	 & $-1.84$	 & $-1.11$	   \\
r\_0017	 & $-0.76$	 & $-$	 & $-0.58$	 & $-0.62$	 & $-$	 & $-0.31$	 & $-0.53$	 & $-0.77$	 & $-0.27$	 & $-0.75$	 & $-0.91$	 & $-0.54$	   \\
r\_0018	 & $-$	 & $-$	 & $-$	 & $-$	 & $-$	 & $-$	 & $-1.65$	 & $-2.48$	 & $-$	 & $-2.30$	 & $-$	 & $-2.01$	   \\
r\_0019	 & $-1.91$	 & $-$	 & $-1.38$	 & $-$	 & $-$	 & $-$	 & $-1.82$	 & $-$	 & $-1.25$	 & $-1.69$	 & $-$	 & $-1.43$	   \\
r\_0020	 & $-2.15$	 & $-$	 & $-2.05$	 & $-1.58$	 & $-1.83$	 & $-1.28$	 & $-2.01$	 & $-$	 & $-1.85$	 & $-1.84$	 & $-1.97$	 & $-1.76$	   \\
r\_0021	 & $-2.03$	 & $-2.18$	 & $-1.96$	 & $-2.31$	 & $-$	 & $-1.97$	 & $-1.95$	 & $-2.14$	 & $-1.81$	 & $-2.15$	 & $-2.31$	 & $-2.07$	   \\
r\_0022	 & $-1.11$	 & $-1.17$	 & $-0.85$	 & $-0.97$	 & $-1.21$	 & $-0.68$	 & $-0.96$	 & $-1.11$	 & $-0.79$	 & $-1.10$	 & $-1.21$	 & $-1.00$	   \\
r\_0024	 & $-0.86$	 & $-0.97$	 & $-0.64$	 & $-0.63$	 & $-0.90$	 & $-0.35$	 & $-0.79$	 & $-1.11$	 & $-0.54$	 & $-1.00$	 & $-1.15$	 & $-0.91$	   \\
b\_0001	 & $-1.99$	 & $-2.11$	 & $-1.66$	 & $-$	 & $-$	 & $-$	 & $-1.62$	 & $-1.93$	 & $-1.30$	 & $-$	 & $-$	 & $-$	   \\
b\_0003	 & $-2.23$	 & $-$	 & $-$	 & $-$	 & $-$	 & $-$	 & $-2.80$	 & $-$	 & $-1.77$	 & $-2.85$	 & $-$	 & $-2.45$	   \\
b\_0005	 & $-0.92$	 & $-$	 & $-$	 & $-1.12$	 & $-$	 & $-$	 & $-$	 & $-$	 & $-$	 & $-1.03$	 & $-$	 & $-$	   \\
b\_0007	 & $-1.55$	 & $-1.62$	 & $-1.34$	 & $-1.51$	 & $-1.70$	 & $-1.25$	 & $-1.21$	 & $-1.38$	 & $-1.08$	 & $-1.35$	 & $-1.49$	 & $-1.26$	   \\
b\_0008	 & $-0.91$	 & $-1.00$	 & $-0.69$	 & $-0.92$	 & $-1.16$	 & $-0.67$	 & $-0.96$	 & $-1.09$	 & $-0.83$	 & $-0.91$	 & $-1.06$	 & $-0.83$	   \\
b\_0009	 & $-1.04$	 & $-$	 & $-$	 & $-1.37$	 & $-$	 & $-$	 & $-1.34$	 & $-$	 & $-$	 & $-$	 & $-$	 & $-$	   \\
b\_0010	 & $-$	 & $-$	 & $-$	 & $-$	 & $-$	 & $-$	 & $-1.49$	 & $-1.97$	 & $-$	 & $-2.16$	 & $-$	 & $-1.84$	   \\
b\_0011	 & $-2.65$	 & $-$	 & $-2.37$	 & $-$	 & $-$	 & $-$	 & $-$	 & $-$	 & $-$	 & $-2.58$	 & $-$	 & $-2.41$	   \\
b\_0015	 & $-$	 & $-$	 & $-$	 & $-$	 & $-$	 & $-$	 & $-2.20$	 & $-$	 & $-2.07$	 & $-1.70$	 & $-1.77$	 & $-1.65$	   \\
b\_0016	 & $-1.83$	 & $-2.00$	 & $-1.78$	 & $-$	 & $-$	 & $-$	 & $-1.99$	 & $-$	 & $-1.91$	 & $-1.68$	 & $-1.81$	 & $-1.62$	   \\
b\_0017	 & $-$	 & $-$	 & $-$	 & $-$	 & $-$	 & $-$	 & $-1.76$	 & $-$	 & $-$	 & $-2.16$	 & $-$	 & $-$	   \\
b\_0018	 & $-$	 & $-$	 & $-$	 & $-$	 & $-$	 & $-$	 & $-$	 & $-$	 & $-$	 & $-$	 & $-$	 & $-$	   \\
b\_0019	 & $-$	 & $-$	 & $-$	 & $-$	 & $-$	 & $-$	 & $-$	 & $-$	 & $-$	 & $-0.99$	 & $-$	 & $-0.51$	   \\
b\_0020	 & $-$	 & $-$	 & $-$	 & $-$	 & $-$	 & $-$	 & $-$	 & $-$	 & $-$	 & $-$	 & $-$	 & $-$	   \\
b\_0023	 & $ 0.11$	 & $-0.19$	 & $ 0.20$	 & $-0.20$	 & $-0.58$	 & $ 0.01$	 & $-0.35$	 & $-0.62$	 & $-0.13$	 & $-0.56$	 & $-$	 & $-0.34$	   \\\hline

\end{tabular}
\tablefoot{\tablefoottext{*}{The star with ID r\_0010 is the target with 99.2\% H4 membership probability.}}
\label{table_a_3}  

\end{sidewaystable*}

\end{appendix}

\end{document}